\documentclass{aa520}
\usepackage{epsfig,color,natbib}
\usepackage[dvips]{rotating}
\usepackage{graphics}
\usepackage{xr,afterpage}

\def\bibfiles{h3317bib}   

\def\aareferences{\longrefs=0  \bibliographystyle{aa}
            \bibliography{h3317aa,\bibfiles}}
\newcount\longrefs
\def\aap{\ifnum\longrefs=1 {Astron.\ Astrophys.}\else 
                           {A\hbox{\rm \&}A}\fi}
\def\aapr{\ifnum\longrefs=1 {Astron.\ Astrophys.\ Rev.}\else 
                            {A\hbox{\rm \&}AR}\fi}
\def\aaps{\ifnum\longrefs=1 {Astron.\ Astrophys.\ Suppl.}\else 
                            {A\hbox{\rm \&}AS}\fi}
\def\aj{\ifnum\longrefs=1 {Astron.\ J.}\else 
                          {AJ}\fi} 
\def\ao{\ifnum\longrefs=1 {Applied Optics}\else 
                           {Appl.\ Opt.}\fi} 
\def\aspcs{\ifnum\longrefs=1 {Astron.\ Soc.\ Pacific Conf. Series}\else 
                           {ASP Conf.\ Ser.}\fi} 
\def\apj{\ifnum\longrefs=1 {Astrophys.\ J.}\else 
                           {ApJ}\fi} 
\def\apjl{\ifnum\longrefs=1 {Astrophys.\ J. Lett.}\else 
                            {ApJ}\fi} 
\def\aplett{\ifnum\longrefs=1 {Astrophys.\ J. Lett.}\else 
                            {ApJ}\fi} 
\def\apjs{\ifnum\longrefs=1 {Astrophys.\ J. Suppl.}\else 
                            {ApJS}\fi}
\def\apss{\ifnum\longrefs=1 {Astrophys.\ and Space Science}\else 
                            {Ap\hbox{\rm \&}SS}\fi}
\def\araa{\ifnum\longrefs=1 {Ann.\ Rev.\ Astron.\ Astrophys.}\else 
                            {ARA\hbox{\rm \&}A}\fi}
\def\azh{\ifnum\longrefs=1 {Astronomicheskii Zhurnal}\else 
                            {Astron.\ Zhur.}\fi}
\def\baas{\ifnum\longrefs=1 {Bull.\ Am.\ Astron.\ Soc.}\else 
                            {BAAS}\fi}
\def\bain{\ifnum\longrefs=1 {Bull.\ Astronom.\ Institutes Netherlands}\else
                            {Bull.\ Astr.\ Inst.\ Neth.}\fi}
\def\gca{\ifnum\longrefs=1 {Geochim.\ Cosmochim.\ Acta}\else 
                           {Geochim.\ Cosmochim.\ Acta}\fi}
\def\grl{\ifnum\longrefs=1 {Geophys.\ Res.\ Lett.}\else 
                           {Geoph.\ Res.\ Lett.}\fi}
\def\iaucirc{\ifnum\longrefs=1 {IAU Circulars}\else 
                          {IAU Circ.}\fi}
\def\ip{\ifnum\longrefs=1 {in press}\else 
                          {in press}\fi}
\def\jgr{\ifnum\longrefs=1 {J.\ Geophys.\ Res.}\else 
                           {J.\ Geophys.\ Res.}\fi}  
\def\jrasc{\ifnum\longrefs=1 {J.\ Royal Astron.\ Soc.\ Canada}\else 
                           {JRAS Can.}\fi}  
\def\mnras{\ifnum\longrefs=1 {Mon.\ Not.\ Roy.\ Astron.\ Soc.}\else 
                             {MNRAS}\fi} 
\def\nat{\ifnum\longrefs=1 {Nature}\else 
                           {Nat}\fi}
\def\pasj{\ifnum\longrefs=1 {Pub.\ Astron.\ Soc.\ Japan}\else 
                            {PASJ}\fi} 
\def\pasp{\ifnum\longrefs=1 {Pub.\ Astron.\ Soc.\ Pacific}\else 
                            {PASP}\fi} 
\def\physscr{\ifnum\longrefs=1 {Physica Scripta}\else 
                            {Phys.\ Scrip.}\fi} 
\def\planss{\ifnum\longrefs=1 {Planetary \& Space Science}\else 
                            {Plan. \& Space Sci.}\fi} 
\def\procspie{\ifnum\longrefs=1 {Proc.\ SPIE}\else 
                            {Proc.\ SPIE}\fi} 
\def\qjras{\ifnum\longrefs=1 {Quarterly J.\ Royal Astron.\ Soc.}\else 
                            {QJRAS}\fi} 
\def\sa{\ifnum\longrefs=1 {Soviet Astron..}\else 
                               {Sov.\ Astron.}\fi}
\def\skytel{\ifnum\longrefs=1 {Sky \& Telescope}\else 
                            {Sky \& Tel.}\fi} 
\def\solphys{\ifnum\longrefs=1 {Solar Phys.}\else 
                               {Solar Phys.}\fi}
\def\ssr{\ifnum\longrefs=1 {Space Science Rev.}\else 
                               {Space\ Sci.\ Rev.}\fi}

\def\dutch{\def\refname{Referenties}\def\abstractname{Samenvatting}%
  \def\bibname{Bibliografie}\def\chaptername{Hoofdstuk}%
  \def\appendixname{Bijlage}\def\contentsname{Inhoudsopgave}%
  \def\listfigurename{Lijst van figuren}\def\listtablename{Lijst van tabellen}%
  \def\indexname{Index}\def\figurename{Figuur}\def\tablename{Tabel}%
  \def\partname{Deel}\def\enclname{Bijlage(n)}\def\ccname{Ter attentie van}%
  \def\headtoname{Aan}\def\headpagename{Pagina}%
  \def\today{\number\day\space\ifcase\month\or januari\or februari\or maart\or%
     april\or mei\or juni\or juli\or augustus\or september\or oktober\or%
     november\or december\fi \space\number\year}%
  \typeout{
              >>>>> use hlatex209 for Dutch hyphenation <<<<< 
         }}
\newcounter{onefig} \newcounter{fignumber}
\newcount\nocaptions \newcount\nofigures \newcount\figwidth
\newcount\viewgraphs
  \def\paper{}  \def\figlabel{} 
\long\def\nextfig#1{\setcounter{figure}{\value{fignumber}}
  \addtocounter{fignumber}{1}
  \ifnum \viewgraphs=1 \newpage \pagestyle{empty} \fi 
  \ifnum\value{onefig}=0 #1 \fi                 
  \ifnum\value{onefig}=\value{fignumber} #1 \fi}
\def\figwidths#1#2{\ifnum \nocaptions=1 #2mm \else #1mm \fi}  
\def\paper#1{}  
\long\def\plotfig#1#2{\ifnum \nofigures=1 \else #2 \fi}
\long\def\captiontext#1{\ifnum \nofigures=1 \raggedright \fi 
   \ifnum \nocaptions=1 \paper
     \ifnum \viewgraphs=0 
       \newline  \mbox{}\hrulefill\mbox{} \newline 
       \newline label:~\{\figlabel\} 
     \fi 
     \else \ifnum \nofigures=0 \fi 
   #1 \fi}

\newcount\panelwidth \newcount\panelheight 
\newcount\bxmin \newcount\bymin \newcount\bxmax \newcount\bymax
\newcount\tbxmin \newcount\tbymin
\newcount\tpanelwidth \newcount\tpanelheight \newcount\tpdif
\def\panelsize #1,#2;{\panelwidth=#1 \panelheight=#2}  
\def\setbb #1,#2;#3,#4;#5,#6;{
  \tbxmin=#1 \tbymin=#2    
  \bxmin=#3 \bymin=#4      
  \bxmax=#5 \bymax=#6}     
\def\barepanel #1{%
  \ifnum\panelheight=0 
    \tpdif=\bymax \advance\tpdif by -\bymin
    \multiply \tpdif by \panelwidth
    \tpanelheight=\tpdif
    \tpdif=\bxmax \advance\tpdif by -\bxmin
    \divide \tpanelheight by \tpdif
  \else \tpanelheight=\panelheight \fi
  \epsfig{file=#1,%
     bbllx=\bxmin bp,bblly=\bymin bp,bburx=\bxmax bp,bbury=\bymax bp,clip=,%
     width=\panelwidth mm,height=\tpanelheight mm}}
\def\labelypanel #1{
  \ifnum\panelheight=0 
    \tpdif=\bymax \advance\tpdif by -\bymin
    \multiply \tpdif by \panelwidth
    \tpanelheight=\tpdif
    \tpdif=\bxmax \advance\tpdif by -\bxmin
    \divide \tpanelheight by \tpdif
  \else \tpanelheight=\panelheight \fi
  \tpdif=\bxmax \advance\tpdif by -\tbxmin
  \tpanelwidth=\panelwidth \multiply \tpanelwidth by \tpdif
  \tpdif=\bxmax \advance\tpdif by -\bxmin
  \divide \tpanelwidth by \tpdif
  \epsfig{file=#1,%
    bbllx=\tbxmin bp,bblly=\bymin bp,bburx=\bxmax bp,bbury=\bymax bp,%
    clip=,width=\tpanelwidth mm,height=\tpanelheight mm}}
\def\labelxpanel #1{%
  \ifnum\panelheight=0 
    \tpdif=\bymax \advance\tpdif by -\bymin
    \multiply \tpdif by \panelwidth
    \tpanelheight=\tpdif
    \tpdif=\bxmax \advance\tpdif by -\bxmin
    \divide \tpanelheight by \tpdif
  \else \tpanelheight=\panelheight \fi
  \tpdif=\bymax \advance\tpdif by -\tbymin
  \multiply \tpanelheight by \tpdif
  \tpdif=\bymax \advance\tpdif by -\bymin
  \divide \tpanelheight by \tpdif
  \epsfig{file=#1,%
    bbllx=\bxmin bp,bblly=\tbymin bp,bburx=\bxmax bp,bbury=\bymax bp,%
    clip=,width=\panelwidth mm,height=\tpanelheight mm}}
\def\labelxypanel #1{%
  \ifnum\panelheight=0 
    \tpdif=\bymax \advance\tpdif by -\bymin
    \multiply \tpdif by \panelwidth
    \tpanelheight=\tpdif
    \tpdif=\bxmax \advance\tpdif by -\bxmin
    \divide \tpanelheight by \tpdif
  \else \tpanelheight=\panelheight \fi
  \tpdif=\bxmax \advance\tpdif by -\tbxmin
  \tpanelwidth=\panelwidth \multiply \tpanelwidth by \tpdif
  \tpdif=\bxmax \advance\tpdif by -\bxmin
  \divide \tpanelwidth by \tpdif 
  \tpdif=\bymax \advance\tpdif by -\tbymin 
  \multiply \tpanelheight by \tpdif
  \tpdif=\bymax \advance\tpdif by -\bymin
  \divide \tpanelheight by \tpdif
  \epsfig{file=#1,%
    bbllx=\tbxmin bp,bblly=\tbymin bp,bburx=\bxmax bp,bbury=\bymax bp,%
    clip=,width=\tpanelwidth mm,height=\tpanelheight mm}}

\def\CC{\par \vspace*{-2ex} \footnotesize \baselineskip=8pt \begin{verbatim}}

\long\def\startignore #1\stopignore{}   

\def\setlistparams{         
  \topsep=0.7ex                 
  \itemsep=0.7ex                
  \leftmargini=3ex}             
\setlistparams                  

\newcounter{alistindex}       

\newcounter{romenumnr}

\newlength{\minipagewidth}

\newsavebox{\boxcontent}
\newcommand{\ovalhead}[1]{
  \unitlength=1cm
  \sbox{\boxcontent}{\mbox{~~{#1}~~}}
  \begin{center}
    \ifdim\wd\boxcontent>6ex 
    \ifdim\wd\boxcontent<8cm 
    \begin{picture}(8,3) \thicklines     
      \put(4.0,0.8){\oval(8,1.6)} 
      \put(0.0,0.7){\parbox{8cm}{
         \begin{center} \usebox{\boxcontent} \end{center}}}
    \end{picture}
    \else \ifdim\wd\boxcontent<12cm 
    \begin{picture}(12,3) \thicklines     
        \put(6.0,0.8){\oval(12,1.6)} 
        \put(0.0,0.7){\parbox{12cm}{
           \begin{center} \usebox{\boxcontent} \end{center}}}
    \end{picture}
    \else
    \begin{picture}(16,3) \thicklines     
        \put(8.0,0.8){\oval(16,1.6)} 
        \put(0.0,0.7){\parbox{16cm}{
           \begin{center} \usebox{\boxcontent} \end{center}}}
    \end{picture}
    \fi \fi \fi
  \end{center}} 

\newcounter{headnr}            
\newcounter{subheadnr}[headnr]
\newcounter{subsubheadnr}[subheadnr]
\def\head #1\par{
  \stepcounter{headnr}                          
  \vspace{2ex} \noindent                        
  {\bf \theheadnr~~~~#1}\\[1ex] \noindent}      
\def\subhead #1\par{  
  \stepcounter{subheadnr}
  \vspace{1.3ex} \noindent
  {\bf \theheadnr.\arabic{subheadnr}~~~#1}\\[0.3ex] \noindent}
\def\subsubhead #1\par{
  \stepcounter{subsubheadnr}
  \vspace{1.0ex} \noindent
  {\bf \theheadnr.\arabic{subheadnr}.\arabic{subsubheadnr}~~~#1}\\ \noindent}

\font\dropfont= cmr12 scaled \magstep5
\def\dropcap#1#2{{\noindent
    \setbox0\hbox{\dropfont #1}\setbox1\hbox{#2}\setbox2\hbox{(}%
    \count0=\ht0\advance\count0 by\dp0\count1\baselineskip
    \advance\count0 by-\ht1\advance\count0by\ht2
    \dimen1=.5ex\advance\count0by\dimen1\divide\count0 by\count1
    \advance\count0 by1\dimen0\wd0
    \advance\dimen0 by.25em\dimen1=\ht0\advance\dimen1 by-\ht1
    \global\hangindent\dimen0\global\hangafter-\count0
    \hskip-\dimen0\setbox0\hbox to\dimen0{\raise-\dimen1\box0\hss}%
    \dp0=0in\ht0=0in\box0}#2}

\def\level #1 #2#3#4{$#1 \: ^{#2} \mbox{#3} ^{#4}$}   

\def\kms{\hbox{km$\;$s$^{-1}$}}

\def\mathstacksym#1#2#3#4#5{\def#1{\mathrel{\hbox to 0pt{\lower 
    #5\hbox{#3}\hss} \raise #4\hbox{#2}}}}

\mathstacksym\lta{$<$}{$\sim$}{1.5pt}{3.5pt} 
\mathstacksym\gta{$>$}{$\sim$}{1.5pt}{3.5pt} 
\mathstacksym\lrarrow{$\leftarrow$}{$\rightarrow$}{2pt}{1pt} 
\mathstacksym\lessgreat{$>$}{$<$}{3pt}{3pt} 

\bibpunct{(}{)}{;}{a}{}{,}
\begin{document}

\newcommand{\ang}{$\rm \AA$}
\newcommand{\tauross}{$\tau_{\mathrm{ross}}$}
\newcommand{\Msun}{M$_{\odot}$}
\newcommand{\Rsun}{R$_{\odot}$}
\newcommand{\Lsun}{L$_{\odot}$}
\newcommand{\be}{\begin{equation}}
\newcommand{\ee}{\end{equation}}
\newcommand{\bee}{\begin{eqnarray}}
\newcommand{\ad}{$\theta_d$}
\newcommand{\vt}{$\xi_t$}
\newcommand{\cc}{$\mathrm{^{12}C/^{13}C}$}
\newcommand{\kn}{$\kappa_{\nu}$}
\newcommand{\lnu}{$l_{\nu}$}
\newcommand{\ha}{H$_{\alpha}$}
\newcommand{\ea}{et al.}
\newcommand{\ene}{\end{eqnarray}}
\newcommand{\teff}{T$_{\mathrm{eff}}$}
\newcommand{\mic}{$\mu$m}
\newcommand{\hoogte}[1]{\rule{0pt}{#1}}
\newcommand{\hminff}{H$^-_{\rm{ff}}$}

\def\offinterlineskip{\baselineskip=-1000pt \lineskip=1pt
\lineskiplimit=\maxdimen}
\def\pra{\mathrel{\mathchoice {\vcenter{\offinterlineskip\halign{\hfil$\displaystyle##$\hfil\cr\propto\cr\sim\cr}}}
{\vcenter{\offinterlineskip\halign{\hfil$\textstyle##$\hfil\cr\propto\cr\sim\cr}}}
{\vcenter{\offinterlineskip\halign{\hfil$\scriptstyle##$\hfil\cr\propto\cr\sim\cr}}}
{\vcenter{\offinterlineskip\halign{\hfil$\scriptscriptstyle##$\hfil\cr\propto\cr\sim\cr}}}}}

\def\ga{\mathrel{\mathchoice {\vcenter{\offinterlineskip\halign{\hfil$\displaystyle##$\hfil\cr>\cr\sim\cr}}}
{\vcenter{\offinterlineskip\halign{\hfil$\textstyle##$\hfil\cr>\cr\sim\cr}}}
{\vcenter{\offinterlineskip\halign{\hfil$\scriptstyle##$\hfil\cr
>\cr\sim\cr}}}
{\vcenter{\offinterlineskip\halign{\hfil$\scriptscriptstyle##$\hfil\cr>\cr\sim\cr}}}}}

\def\la{\mathrel{\mathchoice {\vcenter{\offinterlineskip\halign{\hfil$\displaystyle##$\hfil\cr<\cr\sim\cr}}}
{\vcenter{\offinterlineskip\halign{\hfil$\textstyle##$\hfil\cr<\cr\sim\cr}}}
{\vcenter{\offinterlineskip\halign{\hfil$\scriptstyle##$\hfil\cr
<\cr\sim\cr}}}
{\vcenter{\offinterlineskip\halign{\hfil$\scriptscriptstyle##$\hfil\cr><cr\sim\cr}}}}}


\title{ISO-SWS calibration and the accurate modelling of cool-star atmospheres
\thanks{Based on observations with ISO, an ESA project with
instruments funded by ESA Member States (especially the PI countries France,
Germany, the Netherlands and the United Kingdom) and with the participation of
ISAS and NASA.}}
\subtitle{II.\ General results}

\author{L.~Decin\inst{1}\thanks{\emph{Postdoctoral Fellow of the Fund for
Scientific Research, Flanders}}  \and
B. Vandenbussche\inst{1}\and
C.~Waelkens\inst{1} \and
K.~Eriksson\inst{2} \and
B.~Gustafsson\inst{2} \and
B.~Plez\inst{3} \and
A.J.~Sauval\inst{4} \and
K.~Hinkle\inst{5}
}

\offprints{L.\ Decin, e-mail: Leen.Decin@ster.kuleuven.ac.be}

\institute{Instituut voor Sterrenkunde, KULeuven, Celestijnenlaan 200B, B-3001
    Leuven, Belgium
\and
    Institute for Astronomy and Space Physics, Box 515, S-75120 Uppsala, Sweden
\and
    GRAAL - CC72, Universit\'{e} de Montpellier II, F-34095 Montpellier Cedex 5,
France
\and
    Observatoire Royal de Belgique, Avenue Circulaire 3, B-1180 Bruxelles,
    Belgium
\and
    National Optical Astronomy Observatory\thanks{\emph{Operated by
    the Association of Universities for Research in Astronomy,
    Inc. under cooperative agreement with the National Science
    Foundation}}, P.O. Box 26732, Tucson, Arizona 85726,  USA
}

\date{Received data; accepted date}

\abstract{
The fine calibration of the ISO-SWS detectors (Infrared Space
Observatory - Short Wavelength Spectrometer) has proven to be a
delicate problem. We therefore present a
detailed spectroscopic study in the 2.38 -- 12\,\mic\ wavelength range
of a sample of 16 A0 -- M2 stars used for the calibration of
ISO-SWS. By investigating the discrepancies between the ISO-SWS data
of these sources, the theoretical predictions of their spectra, the
high-resolution FTS-KP (Kitt Peak) spectrum of $\alpha$ Boo and the
solar FTS-ATMOS
(Atmospheric Trace Molecule Spectroscopy) spectrum, both {\it
calibration} problems and problems in {\it
computing the theoretical models and the synthetic spectra} are
revealed. The underlying reasons for these problems are sought for and
the impact on the further calibration of ISO-SWS and on the
theoretical modelling is discussed extensively.
\keywords{Instrumentation: spectrographs -- Methods: data analysis --
Infrared: stars -- Stars: atmospheres -- Stars: late-type -- Stars:
fundamental parameters}}

\maketitle
\markboth{L.\ Decin et al.: ISO-SWS and modelling of cool stars}{L.\ Decin et
al.: ISO-SWS and modelling of cool stars}

\defcitealias{Decin2000A&A...364..137D}{Paper~I}

\section{Introduction}\label{introduction}
For the astronomical community analysing ISO-SWS data
\citep[Infrared Space Observatory, Short-Wavelength
Spectrometer,][]{deGraauw1996A&A...315L..49D}, a first point to
assess when judging and qualifying their data concerns the flux
calibration accuracy. Since the calibration process is not
straightforward, knowledge on the {\it{full}} calibration process
and on the still remaining calibration problems is crucial when
processing the data.

One way to detect calibration problems is by comparing observed data
with theoretical predictions of a
whole sample of standard calibration sources.  But, as explained in
\citet{Decin2000A&A...364..137D} (hereafter referred to as Paper~I) a
full exploitation of the ISO-SWS data may only result from an
iterative process in which both new theoretical developments on the
computation of stellar spectra  --- based on the
MARCS and Turbospectrum code \citep{Gustafsson1975A&A....42..407G,
Plez1992A&A...256..551P, Plez1993ApJ...418..812P}, version May 1998 ---
and more accurate instrumental calibration are involved.

Precisely because this research entails an iterative process, one
has to be extremely careful not to confuse technical detector
problems with astrophysical issues. Therefore, the analysis in its
entirety encloses several steps. Some steps have already been
demonstrated in the case of $\alpha$ Tau in
\citetalias{Decin2000A&A...364..137D}. They will be summarised in
Sect.\ \ref{summary}. Other points will be introduced in Sect.\
\ref{summary} and will be elaborated on in the first sections of
this article (Sect.\ \ref{sample} -- \ref{highres}). Having
described the method of analysis, the general discrepancies
between observed and synthetic spectra are subjected to a careful
scrutiny in order to elucidate their origin.  At this point, a
distinction can be made between discrepancies typically for
{\it{warm}} stars and those typical for {\it{cool}} stars. For
this research, {\it{warm}} stars are defined as being hotter than
the Sun (T$_{\mathrm{eff},\odot} = 5770$~K) and their infrared
spectra are mainly dominated by atomic lines, while molecular
lines are characteristic of cool star spectra. A description on
the general trends in the discrepancies for {\it{warm}} and
{\it{cool}} stars will be made in this paper, while each star of
the sample will be discussed individually in two forthcoming
papers in which also an overview of other published stellar
parameters will be given.

As stated in \citetalias{Decin2000A&A...364..137D}, the detailed spectroscopic
analysis of the ISO-SWS data has till now been restricted to the wavelength
region from 2.38 to 12\,\mic. So, if not specified, the wavelength
range under research is limited to band 1 (2.38 -- 4.08\,\mic) and band 2 (4.08
-- 12.00\,\mic). Band 3 (12.00 -- 29.00\,\mic) will be elaborated on
by Van Malderen (Van Malderen et al., 2001, in prep.).

This paper is organised as follows: in Sect.\ \ref{summary} the
general method of analysis is summarised. The sample of ISO-SWS
observations is described in Sect.\ \ref{sample}, while the data
reduction procedure is discussed in Sect.\ \ref{datareduction}.
The observations of two independent instruments are introduced in
Sect.\ \ref{highres}. In Sect.\ \ref{results}, the results are
elaborated on. In the last section, Sect.\ \ref{impact}, the
impact on the calibration of ISO-SWS and on the theoretical
modelling is given.

The appendix of this article is published electronically. Most of
the grey-scale plots in the article are printed in colour in the
appendix, in order to better distinguish the different spectra.

\section{Description of the strategy}\label{summary}

Since this research includes an iterative process, one has to be very
careful not to introduce (and so to propagate) any errors.
The strategy in its entirety therefore encompasses a number of steps, including
(1.) a spectral coverage of standard infrared sources from A0 to M8,
(2.) a homogeneous data reduction, (3.) a detailed literature study, (4.)
a detailed knowledge of the impact of the various parameters on the
spectral signature, (5.) a statistical method to test the
goodness-of-fit (Kolmogorov-Smirnov test) and (6.) high-resolution
observations with two independent instruments. Some points (in
particular, points 4 and 5) have already been demonstrated in the case
of $\alpha$ Tau in Paper~I.  Points 1, 2 and 6 will be elaborated
on in Sect.\ \ref{sample}, Sect.\ \ref{datareduction} and Sect.\
\ref{highres} respectively.

In its totality, the general method of analysis --- based on these 6
points --- may be summarised as follows:\\
a large set of standard stars (A0 -- M8) has been observed with
ISO-SWS (Sect.\ \ref{sample}). The observational data first have been
subjected to a homogeneous data-reduction procedure (Sect.\
\ref{datareduction}). Thereafter, the carefully reduced ISO-SWS data
of one {\it warm} and one {\it cool} star were compared with
the observational data of two independent instruments (FTS-KP and FTS-ATMOS,
see Sect.\ \ref{highres}). This step is very crucial,
since this is the only secure and decisive way to point out
calibration problems with the
detectors of ISO-SWS. The complete observational data-set, covering a
broad parameter space,  was then
compared with theoretical predictions. By knowing already some
problematic points in the calibration of ISO-SWS, these comparisons
led both to a refinement of our knowledge on the calibration problems
and to a determination of theoretical modelling problems (Sect.\
\ref{results}).

The knowledge on the relative importance of the different molecules
{\footnote{As for atoms, laboratory data for molecules are
generally more accurate than computed data, but much too sparse to
serve as the sole source for the computation of photospheric models or
to represent the myriads of weak lines in a cool star spectrum. An overview
of existing databases can be found in
\citet{Allard1992RMxAA..23..203A, Grevesse1992RMxAA..23...71G,
Gustafsson1994A&ARv...6...19G, Jorgensen1992A&A...261..263J,
Kurucz1992RMxAA..23...45K, Decinthesis}. Some of these, useful for our
research, are described in \citet{Decinthesis}. A copy of this
description can be retrieved from
http://www.ster.kuleuven.ac.be/\~\,$\!$leen. The line lists used for this
research are enumerated in \citetalias{Decin2000A&A...364..137D}.}}
--- displaying their characteristic absorption pattern somewhere in the broad
ISO-SWS wavelength-range --- and on the impact of the various stellar
parameters on the infrared spectrum
enabled us also to determine the fundamental stellar parameters for
the cool giants in our sample \citepalias{Decin2000A&A...364..137D}.
Due to severe calibration problems with the band-2 data (see Sect.\
\ref{datareduction}), only band-1 data were used for this part of the process.
 Once a high-level
of agreement between observed and synthetic data was reached, a
statistical test was needed to objectively judge on the different
synthetic spectra. A choice was made for the Kolmogorov-Smirnov test
\citepalias{Decin2000A&A...364..137D}. This statistical test
{\it{globally}} checks the goodness-of-fit of the observed and
synthetic spectra by computing a deviation estimating parameter
$\beta$ \citepalias[see Eq.\ (5) in][]{Decin2000A&A...364..137D}. The
lower the $\beta$-value, the better the accordance between the
observed data and the synthetic spectrum.

Using this method, the effective temperature, gravity,
metallicity, microturbulent velocity together with the abundance
of C, N and O and the \cc-ratio were estimated for the cool stars.
From the energy distribution of the synthetic spectrum between
2.38 and 4.08\,\mic\ and the absolute flux-values in this
wavelength range of the ISO-SWS spectrum, the angular diameter was
deduced. We therefore have minimised the residual sum of
squares
\begin{equation}
\sum\limits_{i=1}^{n} (f(i) - g(i))^2\,,
\nonumber
\end{equation}
 with $f(i)$  and $g(i)$ representing respectively an
observational and a synthetic data point at the
$i$th wavelength point.

The error bars on the
atmospheric parameters were estimated from 1.\ the intrinsic
uncertainty on the synthetic spectrum (i.e.\ the possibility to
distinguish different synthetic spectra at a specific resolution,
i.e.\ there should be a significant difference in $\beta$-values)
which is thus dependent on both the resolving power of the
observation and the specific values of the fundamental parameters,
2.\ the uncertainty on the ISO-SWS spectrum which is directly
related to the S/N of the ISO-SWS observation, 3.\ the value
of the $\beta$-parameters in the Kolmogorov-Smirnov test and 4.\
the still remaining discrepancies between observed and synthetic
spectra.

It should be noted that an error on the effective temperature
introduces an error on the angular diameter. The IR flux of the {\it cool}
giants does not follow the Rayleigh-Jeans law for a black-body, and we
can write ${\cal{F}}_{\mathrm{obs}}^{\lambda} \propto \mathrm{T_{eff}}^q \cdot
\theta_d^2$. Thus, with $\theta_d \propto
{{\cal{F}}_{\mathrm{obs}}^{\lambda}}^{1/2} \cdot \mathrm{T_{eff}}^{-q/2}$ and
$\sigma_{<{\cal{F}}>} = 0.10 {\cal{F}}$, one obtains
\begin{equation}
\frac{\sigma_{<\theta_d>}}{\theta_d} = \sqrt{\frac{1}{4} (0.10)^2 +
\frac{q^2}{4}
\frac{\sigma^2_{<\mathrm{T_{eff}}>}}{\mathrm{T_{eff}}^2} +
\sigma^2_{\mathrm{int}}}\,,
\label{angdiamq}
\end{equation}
with $\sigma^2_{\mathrm{int}}$ being the intrinsic uncertainty on the angular
diameter (i.e.\ the amount by which one may change the angular
diameter without significant difference in the residual sum of squares).
The uncertainty on the angular diameter (Eq.\ (\ref{angdiamq})) is
mainly determined by the uncertainty in the absolute flux (the first
term in Eq.\ (\ref{angdiamq})), resulting in almost the same
percentage errors in \ad.
Using the ISO-SWS observational data of these
cool giants, we could demonstrate that {\boldmath{$q \approx 1.3 \pm
0.1$ }} for 3600\,K $\le$ \teff $\le$ 4400\,K
\citep{Decinthesis}. With $q \approx
1.3$, $\sigma_{<\mathrm{T_{eff}}>} = 70$\,K and \teff\ = 3850\,K, the
uncertainty increases with a factor 1.008 compared to the $q = 1$
situation.

From the angular diameter and the parallax
measurements (mas) from Hipparcos \citep[with an exception being
$\alpha$ Cen A, for which a more accurate parallax by][is
available]{Pourbaix1999A&A...344..172P}, the stellar radius was
derived. This radius, together
with the gravity --- determined from the ISO-SWS spectrum --- then yielded
the gravity-inferred mass. From the radius and the effective
temperature, the stellar luminosity could be extracted.

This method of analysis could however not be
applied to the {\it{warm}} stars of the sample. Absorption by atoms determines
the spectrum of these stars. It turned out to be unfeasible to determine the
effective temperature, gravity, microturbulence, metallicity and abundances of
the chemical elements from the ISO-SWS spectra of these {\it{warm}}
stars, due to
\begin{itemize}
\item{problems with atomic oscillator strengths in the infrared (see
Sect.\ \ref{hot}),}
\item{the small dependence of the IR continuum on the fundamental
parameters (when changed within their uncertainty),}
\item{the small dependence of the observed atomic line strength on the
fundamental parameters, and}
\item{the absence of molecules, which are each of them specifically dependent on
the various stellar parameters.}
\end{itemize}

Therefore, good-quality published stellar parameters were used to
compute the theoretical model and corresponding synthetic
spectrum. The angular diameter was deduced directly from the ISO-SWS
spectrum, which then yielded --- in conjunction with the assumed parallax,
gravity and effective temperature --- the stellar radius, mass and
luminosity.

\section{Description of the sample}\label{sample}

It seems important to cover a broad parameter space in order to be
able to distinguish between calibration problems and problems
related to the computation of a theoretical model and/or to the
generation of a synthetic spectrum. Therefore,  stellar standard
candles spanning the spectral types A0 -- M8 were observed. The
observations were obtained in the context of two ISO-SWS
proposals. Some calibration data were also provided by the SIDT
(SWS Instrument Dedicated Team) in the framework of a
quick-feedback refining of the model SEDs used for the SWS
calibration. In total, full-resolution SWS scans of  20 standard
stars, covering the full A0 -- M8 spectral range were obtained.
Stars with an earlier spectral type were included in the proposals
of S.\ Price. An overview of the objects, observing dates and
integration times ($t_{\mathrm{int}}$) can be found in Tables
\ref{obs} -- \ref{obs2}. Stars indicated by a `$\bullet$' are
stars which have been scrutinised more carefully than stars
indicated by a `$\circ$'. We first focused on $\alpha$ Bootes
(K2~IIIp), because of its well-known stellar parameters and the
high quality of the ISO-SWS data for this object
\citep{Decin1997fiso.work..185D}. From then on, we have gone, step
by step, towards both higher and lower temperatures. Why the two
warmest stars in our sample (being Vega and Sirius) have been
considered, but have not been studied into all detail, will be
explained in Sect.~\ref{hot}. The other `$\circ$'-stars in Table
\ref{obs2} are stars cooler than an M2 giant --- i.e.\ cooler than
$\sim 3500$~K. Calibration problems with $\gamma$ Cru, variability
\citep{Monnier1998ApJ...502..833M}, the possible presence of a
circumstellar envelope, stellar winds or a warm molecular envelope
above the photosphere \citep{Tsuji1997A&A...320L...1T} made the
use of hydrostatic models for these stars inappropriate and have
led to the decision to postpone the modelling of the coolest stars
in the sample.

\begin{table}[t]
\caption{\label{obs}The logbook of the ISO-SWS
observations  used for this
study. The observing date can be calculated from the revolution number which is
the number of days after 17 November 1995. Sub-bands of bands 1 and 2
observed completely or partially during an AOT06 observation are given
underneath the table. The wavelength ranges for the different sub-bands can be
found in Table \ref{factors}. }
\begin{center}
\setlength{\tabcolsep}{.95mm}
\scriptsize
\begin{tabular}{lrlrclll}\hline
\rule[0mm]{0mm}{5mm} name & HD & HR & HIC & Spectral  &  AOT mode  &
$t_{\mathrm{int}}$ & revo- \\
\rule[-3mm]{0mm}{3mm} & & & & Type & (speed) & [sec] & lution\\
\hline \hline
\rule[0mm]{0mm}{5mm}$\circ$ $\alpha$ Lyr & 172167 & 7001 &
91262 & A0V & AOT01 (3) & 3462 & 178 \\
Vega & & & & & AOT06 & 5642 & 650$^{*1}$ \\
\rule[-3mm]{0mm}{3mm} & & & & & AOT06 & 4354 & 678$^{*2}$ \\
\hline
\rule[0mm]{0mm}{5mm}$\circ$ $\alpha$ CMa & 48915 & 2491 & 32349 & A1V
& AOT01 (4) & 6538 & 689 \\
\rule[-3mm]{0mm}{3mm}Sirius & & & & & AOT01 (1) & 1140 & 868 \\

\hline
\rule[0mm]{0mm}{5mm}$\bullet$ $\beta$ Leo & 102647 & 4534 & 57632 & A3Vv & AOT01
(3) & 3462 & 189\\
\rule[-3mm]{0mm}{3mm}Denebola & & & & & AOT01 (1) & 1096 & 040 \\
\hline
\rule[0mm]{0mm}{5mm}$\bullet$ $\alpha$ Car & & & & & & & \\
\rule[-3mm]{0mm}{3mm}Canopus & {\raisebox{1.5ex}[0pt]{45348}} &
{\raisebox{1.5ex}[0pt]{2326}} & {\raisebox{1.5ex}[0pt]{30438}} &
{\raisebox{1.5ex}[0pt]{F0II}} & {\raisebox{1.5ex}[0pt]{AOT01 (4)}} &
{\raisebox{1.5ex}[0pt]{6538}} & {\raisebox{1.5ex}[0pt]{729}} \\
\hline
\rule[0mm]{0mm}{5mm}$\bullet$ $\alpha$ Cen A & 128620 & 5459 & 71683 & G2V &
AOT01 (4) & 6538 & 607 \\
\rule[-3mm]{0mm}{3mm} & & & & & AOT01 (1) & 1140 & 294 \\
\hline
\rule[0mm]{0mm}{5mm}$\bullet$ $\delta$ Dra & 180711 & 7310 & 94376 &
G9III & AOT1 (4) & 6538 & 206 \\
\rule[-3mm]{0mm}{3mm} & & & & & AOT01 (4) & 6528 & 072 \\
\hline
\rule[0mm]{0mm}{5mm}$\bullet$ $\xi$ Dra & 163588 & 6688 & 87585 & K2III &
AOT01 (3) & 3454 & 314 \\
\rule[-3mm]{0mm}{3mm} & & & & & AOT01 (1) & 1044 & 068 \\
\hline
\rule[-0mm]{0mm}{5mm}$\bullet$ $\alpha$ Boo & 124897 & 5340 & 69673 &
K2IIIp & AOT01 (4) & 6538 & 452 \\
Arcturus  & & & & & AOT01 (4) & 6528 & 071 \\
 & & & & & AOT01 (1) & 1140 & 275 \\
 & & & & & AOT01 (1) & 1094 & 056 \\
& & & & & AOT06 & 3904 & 583$^{*3}$ \\
 & & & & & AOT06 & 4720 & 601$^{*4}$ \\
\rule[-3mm]{0mm}{3mm} & & & & & AOT06 & 4510 & 608$^{*5}$ \\
\hline
\rule[-3mm]{0mm}{8mm}$\bullet$ $\alpha$ Tuc & 211416 & 8502 & 110130 & K3III
& AOT01 (4) & 6539 & 866 \\
\hline \rule[0mm]{0mm}{5mm}$\bullet$ $\beta$ UMi & 131873 & 5563 & 72607 & K4III
& AOT01 (4) & 6546 & 182\\
\rule[-3mm]{0mm}{3mm}Kochab & & & & & AOT01 (2) & 1816 & 079 \\
\hline \hline
\end{tabular}
\end{center}
\footnotesize{ $ ^{*1}$: 1A, 1B, 1D, 1E, 2A, 2B, 2C; $ ^{*2}$: 1A, 1D, 1E, 2A;
$ ^{*3}$: 1A, 1B, 1D, 2A, 2C; $ ^{*4}$: 1A, 1B, 1D, 1E, 2A, 2B, 2C; $ ^{*5}$:
1A, 1B, 1D, 1E, 2A, 2B, 2C}
\end{table}

\begin{table}[t!]
\caption{\label{obs2}Logbook of the ISO-SWS observations: continuation of Table
\ref{obs}.}
\begin{center}
\scriptsize
\setlength{\tabcolsep}{1.mm}
\begin{tabular}{lrrrclll}\hline
\rule[0mm]{0mm}{5mm} name & HD & HR & HIC & Spectral  & AOT
mode & $T_{\mathrm{int}}$ & revo- \\
\rule[-3mm]{0mm}{3mm} & & & & Type & (speed) & [sec] &
 lution\\
\hline \hline \rule[-0mm]{0mm}{5mm}$\bullet$ $\gamma$ Dra & 164058 & 6705 &
87833 & K5III & AOT01 (4) & 6538 & 377 \\
Etamin & & & & & AOT01 (4) & 6542 & 040 \\
 & & & & & AOT01 (2) & 1912 & 811 \\
 & & & & & AOT01 (1) & 1140 & 496 \\
 & & & & & AOT01 (1) & 1062 & 126 \\
 & & & & & AOT06 & 4568 & 501$^{*6}$ \\
\rule[-3mm]{0mm}{3mm}  & & & & & AOT06 & 5676 & 559$^{*7}$ \\
\hline
\rule[-0mm]{0mm}{5mm}$\bullet$ $\alpha$ Tau & 29139 & 1457 & 21421 &
K5III & AOT01(4) & 6538 & 636\\
\rule[-3mm]{0mm}{3mm}Aldebaran & & & & & AOT06$^{*8}$ & 3700 & 681 \\
\hline \rule[-3mm]{0mm}{8mm}$\bullet$ H Sco & 149447 & 6166 & 81304 & K6III &
 AOT01(4) & 6538 & 847 \\
\hline
\rule[0mm]{0mm}{5mm}$\bullet$ $\beta$ And & 6860 & 337 & 5447 & M0III& AOT01 (4)
& 6538 & 795 \\
Mirach & & & & & AOT01 (3) & 3454 & 440 \\
\rule[-3mm]{0mm}{3mm} & & & & & AOT01 (2) & 1912 & 423 \\
\hline
\rule[0mm]{0mm}{5mm}$\bullet$ $\alpha$ Cet & 18884 & 911 & 14135 & M2III& AOT01
(4) & 6538 & 797 \\
\rule[-3mm]{0mm}{3mm}Menkar & & & & & AOT01 (4) & 6538 & 806 \\
\hline
\rule[-0mm]{0mm}{5mm}$\bullet$ $\beta$ Peg & 217906 & 8775 & 113881 &
M2.5III & AOT01(4) & 6538 & 551 \\
Scheat & & & & & AOT01 (3) & 3454 & 206 \\
 & & & & & AOT01 (1) & 1140 & 206 \\
\rule[-3mm]{0mm}{3mm} & & & & & AOT01 (1) & 1096 & 056 \\
\hline
\rule[-0mm]{0mm}{5mm}$\circ$ $\gamma$ Cru & 108903 & 4763 & 61084 & M4III
 & AOT01(4) & 6538 & 609 \\
 & & & & &  AOT01 (2) & 1912 & 258 \\
 & & & & &  AOT01 (2) & 1816 & 079 \\
\rule[-3mm]{0mm}{3mm}  & & & & & AOT06 & 5630 & 643$^{*9}$ \\ \hline
\rule[-3mm]{0mm}{8mm}$\circ$ $\beta$ Gru & 214952 & 8636 & 112122 & M5III
& AOT01 (3) & 3454 &  538 \\
\hline \rule[-3mm]{0mm}{8mm}$\circ$ g Her & 148783 & 6146 & 80704 & M6III
& AOT01 (4) & 6538 & 800 \\
\hline \rule[-3mm]{0mm}{8mm}$\circ$ T Mic & 194676 & & 100935 & M7/8III &
 AOT01 (4) & 6538 & 872 \\ \hline \hline
\end{tabular}
\end{center}
\footnotesize{$ ^{*6}$: 1D, 1E, 2A, 2B, 2C; $ ^{*7}$: 1A, 1B, 1D, 1E, 2A, 2B,
2C; $ ^{*8}$: 1A, 1B, 1D, 1E, 2C; $ ^{*9}$: 1A, 1B, 1D, 1E, 2A, 2B, 2C}
\end{table}
\normalsize

\section{Data reduction}\label{datareduction}

In order to reveal calibration problems, the ISO-SWS data have to
be reduced in a homogeneous way. For all the stars in our sample,
at least one AOT01 observation{\footnote{ Each observation is
determined uniquely by its    
observation number (8 digits), in which the first three digits represent the
revolution number. The observing data can be calculated from the
revolution number which is the number of days after 17 November
1995. }} (AOT\,=\,Astronomical Observation
Template; AOT01\,=\, a single up-down scan for each aperture with
four possible scan speeds at degraded resolution) is available,
some stars have also been observed using the AOT06 mode (\,=\,long
up-down scan at full instrumental resolution). Since these AOT01
observations form a complete and consistent set, they were used as
the basis for the research. In order to check potential
calibration problems, the AOT06 data are used. The scanner speed
of the highest-quality AOT01 observations was 3 or 4, resulting in
a resolving power $\simeq$ 870 or $\simeq$ 1500, respectively
\citep{Leech2002}. The appropriate resolving power of each
sub-band was taken to be the most conservative theoretical
resolving power as determined by Lorente in \citet{Leech2002},
with the exception being band 1A for which this value has been
changed from 1500 to 1300, as will be discussed in Sect.\
\ref{cool}.

The ISO-SWS data were processed to a calibrated spectrum by using the same
procedure as described in \citetalias{Decin2000A&A...364..137D} using
the calibration files available in OLP6.0.

The band 2 (Si:Ga) detectors used in SWS `remember' their
previous illumination history. Going from low to high
illumination, or vice versa, results in detectors asymptotically
reaching their new output value. These are referred to as memory
effects or transients. For sources with fluxes greater than about
100\,Jy, memory effects cause the up and down scans in the SPD
(\,=\,Standard Processed Data) to differ in response by up to
20\,\% in band 2. Since an adapted version of the Fouks-Schubert
model to correct for these memory effects in band 2 was still in
development \citep{Leech2002}, this method could not be applied
during our reduction procedure. Instead, we have used the
down-scan data of our observation as a reference to do a
correction of the flux level of the first scan (up-scan). This is
justified since the memory effects appear to be less severe in the
down-scan measurements, suggesting a more stabilised response to
the flux level for the down-scan data.

Also the band 2 dark current subtraction is closely tied to the
band 2 memory effect correction. The memory effect for Si:Ga detectors
as described by the Fouks-Schubert model is an additive effect. As
such, its proper correction will take place during the dark current
subtraction. Since this correction tool was still not available, all
dark currents were checked individually. When a dark current was
corrupted too much by memory effects, its value was replaced by the
value of a preceding or following dark-current not being affected. In
this way, a small error can occur, which is, however, negligible due
to the high flux level of our stellar sources.

\begin{table}[h!]
\caption{\label{factors} Factors used to multiply the sub-bands for
the highest-quality AOT01 observation (denoted by its revolution
number) of the selected stars in our sample are
given. For each band the wavelength range $\lambda_b$ --
$\lambda_e$ (in \mic) and the (constant) resolving power for a
speed-4 AOT01
observation are also noted. For a speed-3 AOT01 observation one has
to multiply the average resolving power by $\sim 0.58$. In the last two columns
the pointing offset as estimated from known imperfections in the
satellite attitude control system is given in arcsec, where the
y-axis denotes the cross-dispersion direction for SWS and the z-axis the
dispersion direction.}
\begin{center}
\setlength{\tabcolsep}{0.25mm}
\scriptsize
\begin{tabular}{ccccccccrrr}  \hline
\rule[-3mm]{0mm}{8mm} & 1A & 1B & 1D & 1E & 2A & 2B & 2C & rev. & dy & dz \\ \hline \hline
\rule[0mm]{0mm}{5mm}$\lambda / \Delta \lambda$ & 1300 & 1200 & 1500 & 1000 & 1200 & 800 & 800 &
& & \\
$\lambda_b$ [\mic] & 2.38 & 2.60 & 3.02 & 3.52 & 4.08 & 5.30 & 7.00 &
& & \\
\rule[-3mm]{0mm}{3mm}$\lambda_e$ [\mic] & 2.60 & 3.02 & 3.52 & 4.08 & 5.30 &
7.00 & 12.00 & & & \\ \hline
\hline \rule[-3mm]{0mm}{8mm}$\alpha$ Lyr & 1.06 & 1.06  & 1.00 &
1.00 & 1.00 & 0.97 & {\it{+12Jy}} & 178 & $-$0.608  & $-$1.179 \\ \hline
\rule[-3mm]{0mm}{8mm}$\alpha$ CMa & 1.00  & 1.00  & 1.00 & 0.995 &
1.12 & 1.23 &1.00 & 689 & 0.034 & 0.003 \\ \hline
\rule[-3mm]{0mm}{8mm}$\beta$ Leo & 0.99 & 0.99 & 1.00 & 1.00 &
1.16 & 1.27 &{\it{+3.5Jy}} & 189 & 0.478 & 0.556 \\ \hline
\rule[-3mm]{0mm}{8mm}$\alpha$ Car & 0.97 & 0.98 & 1.00 & 1.00 &
0.98 & 1.10 & 0.91 & 729  & 0.024 & 0.072 \\ \hline
\rule[-3mm]{0mm}{8mm}$\alpha$ Cen A & 1.01 & 1.02 & 1.00 & 1.01 &
0.985 & 1.06 & 0.91 & 607  & 0.000 & 0.000 \\ \hline
\rule[-3mm]{0mm}{8mm}$\delta$ Dra & 0.97 & 0.98 & 1.00 & 1.015 &
1.03 & 1.02 & 1.10 & 206 & -0.422 & 1.480 \\ \hline
\rule[-3mm]{0mm}{8mm}$\xi$ Dra & 0.99 & 0.99 & 1.00 & 0.99 & 1.12
& 1.15 & 1.05 & 314 & 1.286 & $-0.282$ \\ \hline
\rule[-3mm]{0mm}{8mm}$\alpha$ Boo & 0.995 & 1.01 & 1.00 & 1.005 &
0.95 & 1.05 & 1.00 & 452  & 0.000 & 0.000 \\ \hline
\rule[-3mm]{0mm}{8mm}$\alpha$ Tuc & 1.005 & 1.02 & 1.00 & 1.01 &
1.05 & 1.00 & 1.00 & 866  & 0.000 & 0.000 \\ \hline
\rule[-3mm]{0mm}{8mm}$\beta$ UMi & 1.00 & 1.015 & 1.00 & 1.01 &
0.91 & 0.885 & 1.00 & 182 & $-1.062$ & 0.045 \\ \hline
\rule[-3mm]{0mm}{8mm}$\gamma$ Dra & 0.995 & 1.005 & 1.00 & 1.005 &
0.935 & 0.98 & 0.91 & 377 & $-0.304$ & 0.181 \\ \hline
\rule[-3mm]{0mm}{8mm}$\alpha$ Tau & 1.00 & 1.01 & 1.00 & 1.00 &
1.00 & 1.045 & 1.00 & 636 & 0.000 & 0.000 \\ \hline
\rule[-3mm]{0mm}{8mm}H Sco & 1.00 & 1.015 & 1.00 & 1.00 & 1.13
& 1.05 & 1.15 & 847 & 0.000 & 0.000 \\ \hline
\rule[-3mm]{0mm}{8mm}$\beta$ And & 1.00 & 1.00 & 1.00 & 1.005 &
1.00 & 1.10  & 0.95 & 795 & 0.000 & 0.000 \\ \hline
\rule[-3mm]{0mm}{8mm}$\alpha$ Cet & 0.985 & 1.00 & 1.00 & 1.01 &
0.935 & 1.03 & 0.91 & 797& 0.000 & 0.000 \\ \hline
\rule[-3mm]{0mm}{8mm}$\beta$ Peg & 1.00 & 1.015 & 1.00 & 1.005 &
0.935 & 1.03 & 0.935 & 551 & 0.017 & 0.095 \\ \hline
\end{tabular}
\end{center}
\end{table}

Since the different sub-band spectra can show jumps in flux at the
band-edges, several sub-bands had to be multiplied by a small
factor to construct a smooth spectrum.  Three causes for the
observed shift factors between different sub-bands of an
observation and between different observations of a given stellar
source can be reported: 1.\ pointing errors, 2.\ problems with the
RSRF correction, and 3.\ a problematic dark current subtraction,
from which the pointing errors are believed to have the largest
impact. The pointing errors as well as the RSRF correction causes
a decrease in flux by a gain factor, while the dark current
subtraction can lower the flux level by an offset. As the effects
of the pointing errors are estimated to have the biggest effect,
and since the stars in our sample have a high flux level so that
the dark current subtraction only plays a marginal role, the
individual sub-bands were multiplied with a factor --- rather than
shifted with an offset --- in order to obtain a smooth spectrum.
These factors (see Table \ref{factors}) were determined by using
the overlap regions of the different sub-bands and by studying the
other SWS observations.  The band-1D data were taken as reference
data, due to the absence of strong molecular absorption in
this wavelength range which may cause a higher standard deviation
in the bins obtained when rebinning the oversampled spectrum, and
--- most importantly --- due to the low systematic errors in this
band, caused by e.g.\ errors in the curve of the RSRF, detector
noise, uncertainties in the conversion factors from $\mu$V/s to
Jy, ... \citep{Leech2002}. Using the total absolute uncertainty
values --- which have accumulated factors from each of the
calibration steps plus estimated contributions from processes
which were unprobed or uncorrected --- as given in Table 5.3 in
\citet{Leech2002}, the estimated 1\,$\sigma$ uncertainty on these
factors is 10\,\%. As is clearly visible from Table
\ref{factors}, these factors do not show any trend with spectral
type or flux-level. This is displayed in Fig.\
\ref{ratiosfig}, where the band-border ratios between 1A-1B, 1B-1D
and 1D-1E are plotted in function of the flux at 2.60, 3.02 and
3.52 $\mu$m respectively. For this plot, all the observations of
the cool stars in our sample, discussed in the Appendix of
Paper~IV of this series, are used. In band 1, the band-border
ratios of 1A-1B and 1D-1E are from bands within the same aperture.
Going from band 1B to band 1D, the aperture changes. Satellite
mispointings can have a pernicious impact on this band-border
ratio: the mean deviation of the band-border ratios w.r.t.\ 1 is
significantly larger for 1B-1D ($=0.015$) than for 1A-1B and 1D-1E
(being respectively 0.009 and 0.005). Due to the problems with
memory effects in band 2 (4.08 -- 12\,\mic), the factors of each
sub-band of band 2 were determined by use of the corresponding
spectral data of Cohen \citep{Cohen1992AJ....104.2030C,
Cohen1995AJ....110..275C, Cohen1996AJ....112.2274C,
Witteborn1999AJ....117.2552W}: for Vega and Sirius Cohen has
constructed a calibrated model spectrum; a composite spectrum
(i.e.\ various observed spectra have been spliced to each other
using photometric data) is available for $\alpha$ Cen A, $\alpha$
Boo, $\gamma$ Dra, $\alpha$ Tau, $\beta$ And, $\alpha$ Cet, and
$\beta$ Peg; a template spectrum (i.e.\ a spectrum made by using
photometric data of the star itself and the shape of a `template'
star) is built for $\delta$ Dra (template: $\beta$ Gem: K0~III),
$\xi$ Dra (template: $\alpha$ Boo: K2~IIIp), $\alpha$ Tuc
(template: $\alpha$ Hya: K3~II-III) and H Sco (template: $\alpha$
Tau: K5~III). When no template was available (for $\beta$
Leo, $\alpha$ Car and $\beta$ UMi), the synthetic spectrum
showing the best agreement with the band-1 data was used as
reference. This does not imply that we are trapped in a circular
argument, since the stellar parameters for the synthetic spectrum
were determined from the band-1 data only. Moreover, the
maximum difference in the correction factors for band 2 obtained
when using the synthetic spectra instead of a Cohen template for
the 13 stars common in the sample is 7\,\%, which is well within
the photometric absolute flux uncertainties claimed by
\citet{Leech2002}. Note that all shift factors are in within the
AOT01 band border ratios as derived in Fig.\ 5.33 and Fig.\ 5.34
in \citet{Leech2002}. Using the overlap regions in band 2 can
have quite a big effect on the final composed spectrum: focussing
on $\beta$ UMi, we note that by using these overlap regions band
2A (and consequently bands 2B and 2C) should be shifted downwards
by a factor 1.04; in order then to match the shifted band 2B and
band 2C, band 2C should be once more shifted downwards by a factor
1.12. In general, the error in the absolute flux could increase to
$\sim 20$\,\% at the end of band 2C when this method would be
used.

For a more elaborate discussion on the SWS error budget, we would like
to refer to \citet{Leech2002}.

\begin{figure}[h!]
\begin{center}
\resizebox{0.5\textwidth}{!}{\rotatebox{0}{\includegraphics{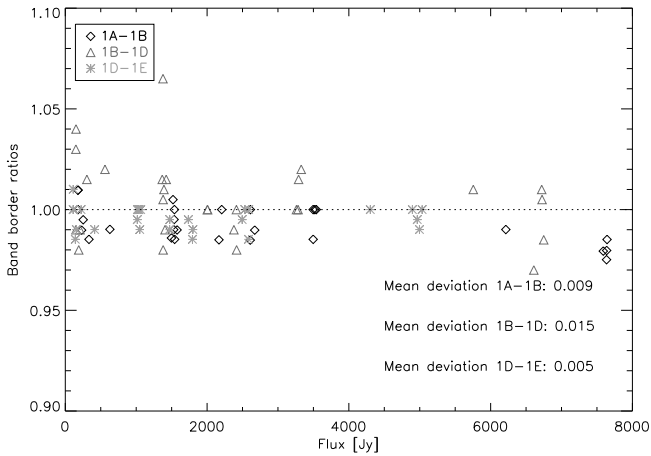}}} 
\caption{\label{ratiosfig}Flux ratios between the different sub-bands
of band 1 at the wavelengths of overlap. The mean deviation w.r.t.\ 1
is given for the different band borders at the right corner of the figure.}
\end{center}
\end{figure}

\section{Verification of the results with FTS-KP and FTS-ATMOS
spectra}\label{highres}

To test our findings --- indicating either a calibration problem or a
theoretical modelling problem --- with data taken with an independent
instrument, a high-resolution observation of both one {\it{warm}}
and one {\it{cool}} star were included. The
high-resolution Fourier Transform Spectrometer (FTS) spectrum of
$\alpha$ Boo \citep{Hinkle1995PASP..107.1042H} and the Atmospheric Trace
Molecule Spectroscopy (ATMOS) spectrum of the
Sun \citep{Farmer1989QB551.F37......, Geller1992} are used as external
control to the process.

The Arcturus observations were obtained with the FTS at the Kitt
Peak (KP) 4m Mayall telescope mainly in 1993 and 1994 and are described in
\citet{Hinkle1995, Hinkle1995PASP..107.1042H}. The entirety of the 1 to
5 $\mu$m Arcturus
spectrum detectable from the ground was observed twice at opposite
heliocentric velocities, allowing the removal of many telluric
lines. We refer to these two
spectra as the winter and summer observations.  The data were obtained
at a resolving power ($\lambda$/$\Delta\lambda$) near 100000.
Spectra of the Earth's atmosphere, derived from high resolution solar
spectra, have been ratioed to the Arcturus spectra to largely remove the
telluric lines.  Most telluric lines less than 30\,\% deep are cleanly
removed but as the optical depth of the telluric lines increases more
information is lost and lines over 50\,\% deep can not be removed.  Gaps
in the plots appear at these spectral regions.  A typical example of
the Arcturus spectra is shown in Fig.\ \ref{FTS} on page
\pageref{FTS}.  In the upper panel of Fig.\
\ref{FTS}, some CO lines of the first overtone are plotted.  In the lower
panel of Fig.\ \ref{FTS}, one can clearly distinguish atomic and OH lines.


During the period from November 3 to November 14, 1994, the Atmospheric Trace
Molecule Spectroscopy (ATMOS) experiment operated as part of the
ATLAS-3 payload of the shuttle Atlantis
\citep{Gunson1996GeoRL..23.2333G}. The principal
purpose of this experiment was to study the distribution of the
atmosphere's trace molecular components. The
instrument, a modified Michelson interferometer, covering the
frequency range from 625 to 4800\,cm$^{-1}$ at a spectral
resolution of 0.01\,cm$^{-1}$, also recorded
high-resolution infrared spectra of the Sun. A small part of the
intensity spectrum is shown in
Fig.\ \ref{sun}. The Holweger-M\"{u}ller model
\citep{HolwegerMuller1974SoPh...39...19H} was used as input
for the computation of the synthetic spectrum of the Sun. A
microturbulent velocity of 1\,\kms\ was assumed.

\begin{figure}[h!]
\begin{center}
\resizebox{0.5\textwidth}{!}{\rotatebox{90}{\includegraphics{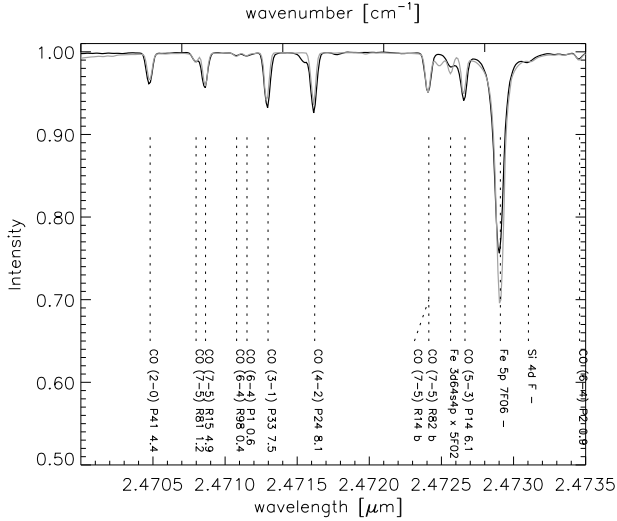}}}
\caption{\label{sun} FTS-ATMOS intensity spectrum of the Sun (black)
at a resolving power $\lambda / \Delta \lambda$ of 60000. The
synthetic spectrum (grey) has been computed using the
Holweger-M\"{u}ller model \citep{HolwegerMuller1974SoPh...39...19H} with \vt\ =
1\,\kms.}
\end{center}
\end{figure}

In the case of $\alpha$ Boo, a high-resolution synthetic spectrum was first
generated with the stellar parameters of
\citet{Kjaergaard1982A&A...115..145K} as input parameters (\teff\ =
4350\,K, $\log$ g = 1.80, [Fe/H] = $-0.50$, $\varepsilon$(C) = 7.89,
$\varepsilon$(N) = 7.61, $\varepsilon$(O) = 8.68). For the
microturbulence \vt\  a
velocity of 1.7\,\kms\ was assumed as being the mean for red giants
\citep{Gustafsson1974A&A....34...99G}. Since both
\citet{Mackle1975A&AS...19..303M} and \citet{Bell1985MNRAS.212..497B}
found a Si and Mg abundance depleted with respect to the solar values,
$\varepsilon$(Mg) was taken to be 7.33 and $\varepsilon$(Si) to be
7.20. For
the anisotropic macroturbulence $\Gamma_t$, a radial-tangential
profile was assumed \citep{Gray1975ApJ...202..148G} with FWHM of 3\,\kms.
This synthetic spectrum was then convolved with the beam-profile
function (i.e.\ a sinc-function). Both observed FTS-KP and synthetic
spectrum were
rebinned to a resolving power of 60000, since the resolving power of
the FTS-KP was not constant over the whole wavelength range.
Using this set of input-parameters, the high-excitation CO ($\Delta v = 1$ and
$\Delta v = 2$) and SiO ($\Delta v = 2$) lines were predicted as being
too weak. This was confirmed from both the FTS-KP and the
ISO-SWS spectrum. Especially from the ISO-SWS spectrum --- and more
specifically from the slope in bands 1B and 1D --- it became clear that
the gravity should be lowered. Using then
the FTS-KP-spectra, the input parameters were
improved until an optimal fit was obtained (see Fig.\ \ref{FTS} on
page \pageref{FTS}). This
resulted in the
following parameters: \teff\ = 4350\,K, $\log$ g =
1.50, [Fe/H] = $-0.50$, \vt\ = 1.7\,\kms, \cc\ = 7,
$\varepsilon$(C) = 7.96, $\varepsilon$(N) = 7.61, $\varepsilon$(O)
= 8.68, $\varepsilon$(Mg) = 7.33 and $\varepsilon$(Si) = 7.20.

Discrepancies appearing from the confrontation of the synthetic
spectrum with both the FTS-KP and the ISO-SWS spectra,
may be clearly
attributed to problems in constructing the model or generating the
synthetic spectrum. Other discrepancies in the
ISO-SWS versus synthetic spectrum are the ones caused by calibration
problems.

For the {\it{warm}} stars in the sample, an analogous
comparison was performed with the help of the FTS-ATMOS intensity
spectrum of the Sun. Discrepancies between the ISO-SWS and
synthetic spectra of $\alpha$ Cen A (G2V) were compared with the
discrepancies found in the comparison between the FTS-ATMOS spectrum
of the Sun and its synthetic spectrum (Fig.\ \ref{sun}).

Inspecting Fig.\ \ref{sun} and Fig.\ \ref{FTS}, some discrepancies become
immediately visible, e.g.\ the CO 2-0 and 3-1 are predicted too weak
in both figures. The reason for this and other discrepancies will be
explained in the next section.

\section{Results}\label{results}


Computing synthetic spectra is one step, distilling useful
information from it is a second --- and far more difficult --- one.
Fundamental stellar parameters for this sample of bright stars are
a first direct result which can be deduced from this comparison
between ISO-SWS data and synthetic spectra. In papers III and IV of this series,
these parameters will be discussed and confronted with other published stellar
parameters.

\begin{figure}[h!]
\begin{center}
\resizebox{0.5\textwidth}{!}{\rotatebox{90}{\includegraphics{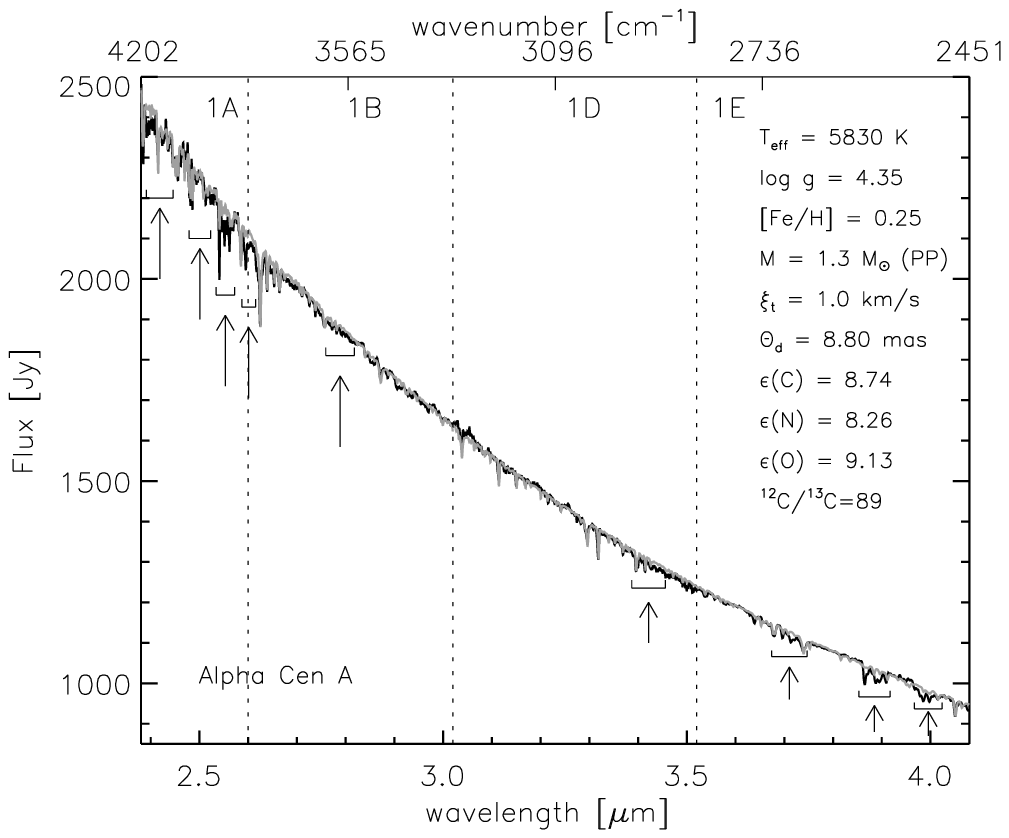}}}
\caption{\label{acenarrow} Comparison between the ISO-SWS data of
$\alpha$ Cen A (black) and the synthetic spectrum (grey) with
stellar parameters \teff\ = 5830\,K, $\log$ g = 4.35, M = 1.3\,\Msun,
[Fe/H] = 0.25,
\vt\ = 1.0\,\kms, \cc\ = 89, $\varepsilon$(C) =
8.74, $\varepsilon$(N) = 8.26, $\varepsilon$(O) = 9.13 and \ad\ =
8.80\,mas. Some of the most prominent discrepancies between these
two spectra are indicated by an arrow.
A coloured version of this plot is available in the Appendix as Fig.\
\ref{acenarrowcol}.}
\end{center}
\end{figure}

\begin{figure}[h]
\begin{center}
\resizebox{0.5\textwidth}{!}{\rotatebox{90}{\includegraphics{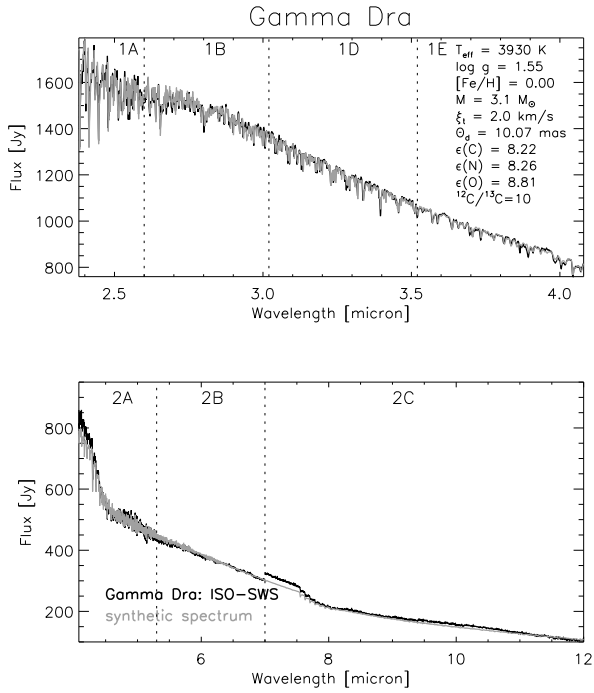}}}
\caption{\label{gamdra} Comparison between band 1 and band 2 of
the ISO-SWS data of $\gamma$ Dra (black) and the synthetic
spectrum (grey) with stellar parameters \teff\ = 3960\,K, $\log$ g =
1.30, M = 1.7\,\Msun, [Fe/H] = 0.00, \vt\ = 2.0\,\kms, \cc\ = 10,
$\varepsilon$(C) = 8.15, $\varepsilon$(N) = 8.26, $\varepsilon$(O)
= 8.93 and \ad\ = 9.98\,mas. A coloured version of this plot is
available in the Appendix as Fig.\ \ref{gamdracol}.}
\end{center}
\end{figure}

A typical example of both a {\it{warm}} and {\it{cool}} star is given in
Fig. \ref{acenarrow} and Fig. \ref{gamdra} respectively. Different types of
discrepancies do emerge. The size of the discrepancies between
ISO-SWS observations and theoretical predictions varies a lot.
Both for the {\it{warm}} and for the {\it{cool}} stars we see a general
good agreement in shape between the ISO-SWS and theoretical data in
band 1, with however (local) error peaks up to $\sim 8\,\%$.  An exception
is $\beta$ Peg for which the shape is wrong by $\sim 6\,\%$ and local
error peaks may go up to $\sim 15\,\%$. The agreement between
observational and synthetic data is worse in band 2: a general
mismatch by up to $\sim 15\,\%$ may occur.

By scrutinising carefully the various discrepancies between the
ISO-SWS data and the synthetic spectra of the standard stars in
our sample, the origin of the different discrepancies was
elucidated. First of all, a description on the general trends in
discrepancies for the {\it{warm}} stars will be made, after which
the {\it{cool}} stars will be discussed.

\subsection{Warm stars: A0 -- G2}\label{hot}

\begin{enumerate}

\item{Especially when concentrating on $\alpha$ Cen A (G2~V),
one notifies quite a few spectral features which appear in the
ISO-SWS spectrum, but are absent in the synthetic spectrum. By
comparing the spectra of the other warm stars ($\beta$ Leo,
$\alpha$ Car, $\alpha$ Lyr and $\alpha$ CMa) with each other,
corresponding spectral features can be recognised in their ISO-SWS
spectra, although this is somewhat more difficult for $\beta$ Leo
due to the lower resolving power and lower signal-to-noise ratio.
Also for the cooler stars in the sample, these spectral features
are (weakly) present. Some of the most prominent ones are
indicated by arrows in the spectrum of $\alpha$ Cen A in Fig.\
\ref{acenarrow}. Molecular absorption can not be a possible
cause of/ contribution to e.g.\ the 3.9 $\mu$m feature, since this
feature is seen for the warm stars in the sample, where no
molecular absorption occurs at these wavelength ranges. The
solar FTS-ATMOS spectrum proved to be extremely useful for the
determination of the origin of these features. All spectral
features, indicated by an arrow in Fig.\ \ref{acenarrow}, turned
out to be caused by --- strong --- atomic lines (Mg, Si, Fe, Al,
C, ...). This is illustrated for the wavelength range from 3.85 to
3.92 \mic\  in Fig.\ \ref{atomic}. In panel (a) of Fig.\
\ref{atomic}, the high-resolution FTS-ATMOS spectrum of the Sun is
compared with its synthetic spectrum in the wavelength range from
3.853 to 3.917 \mic. The strongest lines are identified by using
the line list of \citet{Geller1992}. At the ISO resolving power of
1000 in band 1E, these atomic lines are reduced to the features
indicated in panel (b). The same features can be recognised in the
ISO-SWS spectrum of $\alpha$ Cen A in panel (c). It is clear that
these atomic features are not well calculated for the synthetic
spectra of the Sun and $\alpha$ Cen A.

\begin{figure}[h!]
\begin{center}
\resizebox{0.5\textwidth}{!}{\rotatebox{90}{\includegraphics{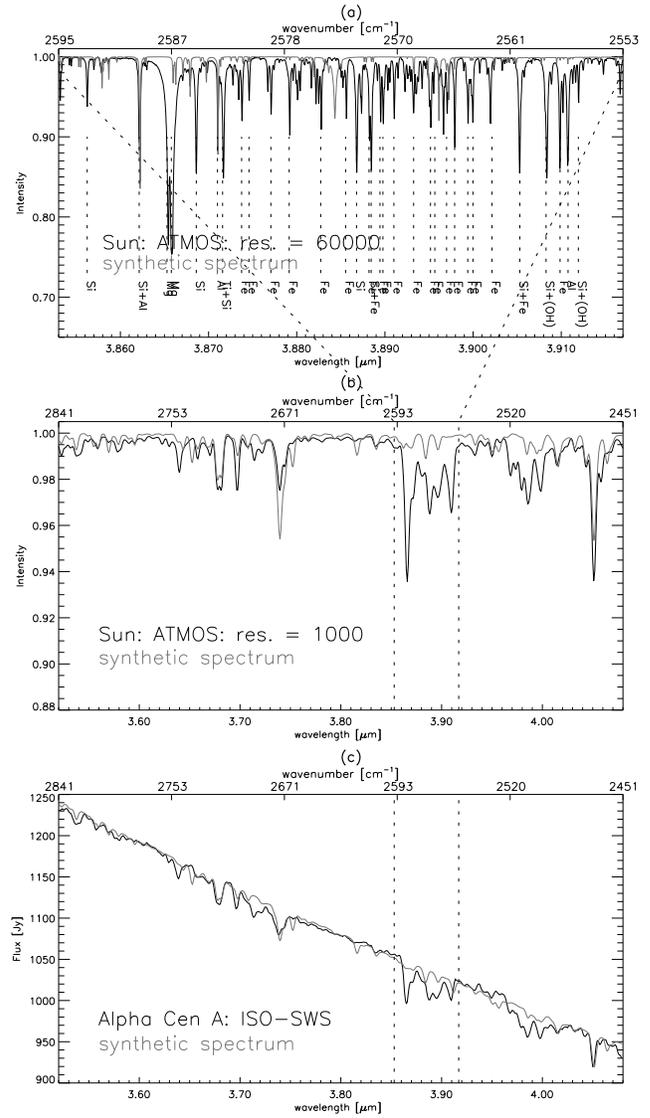}}}
\caption{\label{atomic}In panel (a) the high-resolution FTS-ATMOS
spectrum of the Sun is compared with its synthetic spectrum based
on the Holweger-M\"{u}ller model (1974) and computed by using the
atomic line list of \citet{Hirhor95}. Panel (b) shows the same
comparison as panel (a), but now at a resolving power of 1000 for
the wavelength range going from 3.52 -- 4.08\,\mic\ (band 1E). The
data of band 1E of the ISO-SWS spectrum of $\alpha$ Cen A (at a
resolving power of 1000) are plotted in panel (c).}
\end{center}
\end{figure}

For Fig.\ \ref{atomic}, the atomic line list of \citet{Hirhor95} was used to
generate the synthetic spectrum. This line list has as starting files the
compilation by \citet{Kurucz1975SAOSR.362.....K} and \citet{Kurucz1989}
who lists the semi-empirical $gf$-values for many
ions. Energy levels were adopted from recent compilations
\citep{Sugar1985aeli.book.....S} or individual works. The line list of
\citet{Kelly1983} was merged into the file. The $gf$-values are taken from
several compilations \citep[e.g.][]{Fuhr1988atpi.book.....F,
Fuhr1988atps.book.....F, Reader1980wtpa.book.....R, Wiese1966atp..book.....W,
Wiese1969atp..book.....W, Morton1991ApJS...77..119M, Morton1992ApJS...81..883M}.
Published results of the Opacity Project \citep{Seaton1995QB809.O63......} were
also included. Comparing the atomic line list of \citet{Hirhor95} with
the identifications as given by \citet{Geller1992} made clear that
quite some lines are misidentified or not included in the IR atomic line list of
\citet{Hirhor95}.

The usage of VALD \citep[Vienna Atomic Line
Database:][]{Piskunov1995lahr.conf..610P, Ryabchikova1997BaltA...6..244R,
Kupka1999A&AS..138..119K}{\footnote{No more recent ``near IR'' data for neutral metal lines have
been added to VALD, so no improvements for this particular wavelength
range could be expected over the \citet{Kurucz1992RMxAA..23...45K}
line lists.}} and the line list of \citet{Vanhoof1998sese.conf...67V}
did not solve the problem.
Obviously, the oscillator strengths of the atomic lines in the
infrared are not known sufficiently well.

In order to test this hypothesis, Sauval (2002, {\it{priv.\
comm.}}) has constructed a new atomic line list by deducing new
oscillator strengths from the high-resolution ATMOS spectra of the
Sun (625 -- 4800 cm$^{-1}$). A preliminary comparison using this
new linelist is given in Fig.\ \ref{atomicsauv} (which should be
compared with Fig.\ \ref{atomic}). The contents, nature,
limitations, uncertainties, \ldots \ of this new line list will be
discussed in a forthcoming paper. But it is already
obvious from a confrontation between Fig.\ \ref{atomic} and Fig.\
\ref{atomicsauv} that these new oscillator strengths from Sauval
are more accurate.

\begin{figure}[h!]
\begin{center}
\resizebox{0.5\textwidth}{!}{\rotatebox{90}{\includegraphics{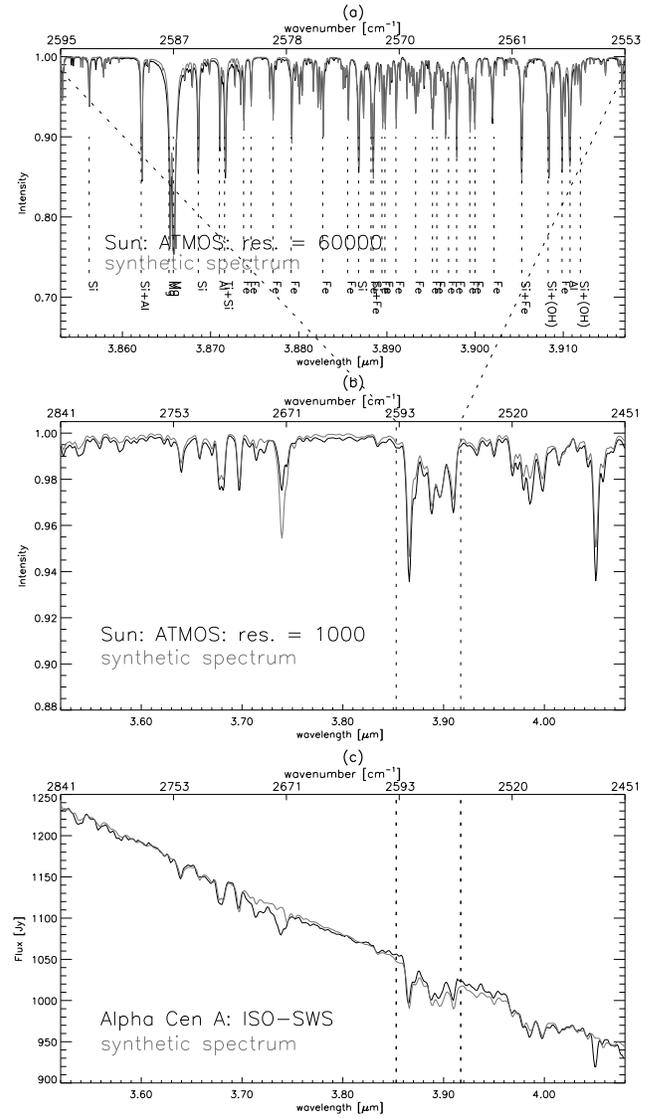}}}
\caption{\label{atomicsauv}In panel (a) the high-resolution ATMOS
spectrum of the Sun is compared with its synthetic spectrum based
on the Holweger-M\"{u}ller model (1974) and computed by using the
atomic line list of Sauval (2002, {\it{priv.\ comm.}}). Panel (b) shows the same
comparison as panel (a), but now at a resolving power of 1000 for
the wavelength range going from 3.52 -- 4.08\,\mic\ (band 1E). The
data of band 1E of the ISO-SWS spectrum of $\alpha$ Cen A (at a
resolving power of 1000) are plotted in panel (c).}
\end{center}
\end{figure}

It is
plausible that the same reason, i.e.\ wrong and missing
oscillator strengths of atomic lines in the infrared, together with
noise, is the origin of the observed discrepancies between the
ISO-SWS and synthetic spectra for the other warm stars in the sample, because
\begin{itemize}
\item{incomplete atomic data already caused a problem for the
synthetic spectrum of $\alpha$ Cen A, and}
\item{the stars concerned are hotter than $\alpha$ Cen A, so
atoms (maybe of higher excitation or ionisation) will determine
even more the spectral signature of these stars.}
\end{itemize}
The lack of reliable atomic data rendered the determination of the fundamental
stellar parameters for the {\it{warm}} stars from the ISO-SWS data
impossible. Another consequence concerns the continuum, which was very difficult
to determine. Therefore,  the uncertainty on the angular diameter of
these {\it{warm}} stars is more pronounced.}

\item{Secondly, the hydrogen lines are conspicuous. For example,
the synthetic hydrogen Pfund lines are almost always
predicted as being too strong for main-sequence stars, while they are
predicted as being too weak for the supergiant $\alpha$ Car.
This indicates a problem with the generation of the synthetic
hydrogen lines, which is corroborated when the high-resolution
FTS-ATMOS spectrum of the Sun is compared with its synthetic spectrum
(see Fig.\ \ref{sunHlines}). In the TurboSpectrum program
\citep{Plez1992A&A...256..551P,Plez1993ApJ...418..812P}  the
hydrogen line-profile calculation, adopted from the SYNTHE code of
Kurucz, includes the Stark, the Doppler,
the van der Waals and the resonance broadening. For the first four
lines in every series, the fine structure is also included for the
core calculations. The problematic computation of the self broadening of
hydrogen lines \citep{Barklem2000A&A...355L...5B} and the unproper treatment of
the convection by using the mixing-length theory (see, e.g.,
\citet{Asplund2000A&A...359..729A} where realistic ab-initio 3D
radiative-hydrodynamical convection simulations have been used) may account for
part of this discrepancy.

\begin{figure}[h!]
\begin{center}
\resizebox{0.5\textwidth}{!}{\rotatebox{90}{\includegraphics{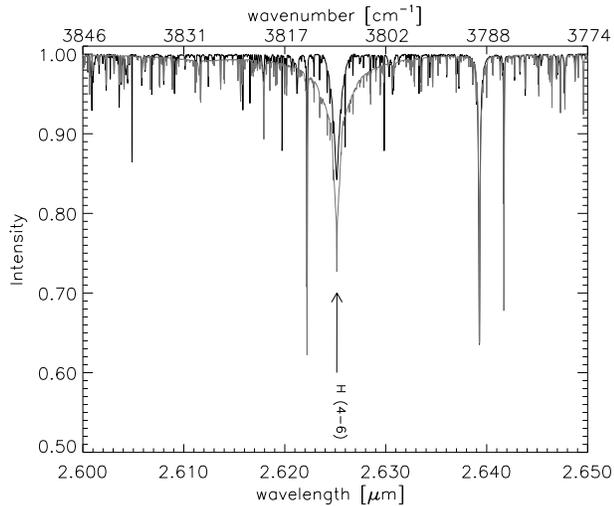}}}
\caption{\label{sunHlines} FTS-ATMOS intensity spectrum of the Sun
(black) at a resolving power $\lambda / \Delta \lambda$ of 60000.
The synthetic spectrum (grey) has been computed using the
Holweger-M\"{u}ller model \citep{HolwegerMuller1974SoPh...39...19H} with \vt\ =
1\,\kms. The Bracket
$\beta$ line is indicated by an arrow. Most of the other features
are CO $\Delta v = 2$ lines, some are atomic lines.}
\end{center}
\end{figure}
}

\item{Compared to the ISO-SWS data, the synthetic spectra of {\it{warm}} stars
display a higher synthetic flux between the H5-9 and H5-8 hydrogen
line (see Fig.\ \ref{humphreys}). From other SWS observations
available in the ISO data-archive, we could deduce that this
`pseudo-continuum' starts arising for stars hotter than K2 ($\sim
4500$\,K). Since this effect is {\it{systematically}} seen
for {\it{all}} {\it{warm}} stars in the sample, the origin of this
discrepancy can not be a wrong multiplication factor. Note that
band 1D and band 1E overlap each other over quite a large
wavelength range, so that the multiplication factor of band 1E can
be accurately determined. Moreover such an effect was not seen
for the cooler K and M giants, the spectrum of which is dominated
by OH features in this wavelength range, so we could reduce the
problem as having an atomic origin. A scrutiny on the hydrogen
lines shows that the high-excitation Humphreys-lines (from H6-18
on) --- and Pfund-lines --- are always calculated as too weak.
Moreover, the Humphreys ionisation edge occurs at 3.2823\,\mic.
Since the discrepancy does not appear above the limit (i.e.\ at
shorter wavelengths) and is disappearing beyond the
Brackett-$\alpha$ line, the conclusion is reached that a
problem with the computation of the complex line profile of the
crowded Humphreys hydrogen lines towards the series limit causes
this discrepancy. The higher the transition, i.e.\ going from
H6-18 to H6-93, the more the $\log$ gf-value should be increased
in order to get a good match between observations and theoretical
computations: increasing $\log$ gf with $+0.3$\,dex gives us a
good match for the H6-18 line in $\alpha$ Car, while we should
increase $\log$ gf with $+1.0$\,dex for the H6-16 line in $\alpha$
Car. We may conclude that a complex problem with the computation
of the pressure broadening will be on the origin of the discussed
discrepancy.

\begin{figure}[h!]
\begin{center}
\resizebox{0.5\textwidth}{!}{\rotatebox{90}{\includegraphics{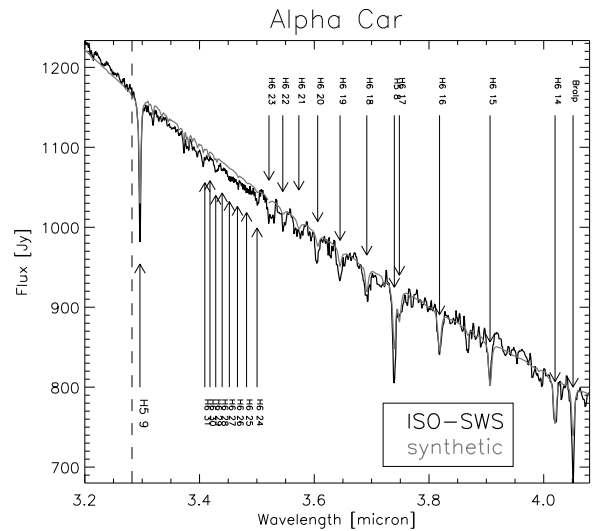}}}
\caption{\label{humphreys} Comparison between the observed ISO-SWS
spectrum of $\alpha$ Car and its synthetic spectrum in the
wavelength region from 3.20 to 4.08\,\mic. The Brackett-$\alpha$,
two Pfund-lines (H5-8 and H5-9) and a number of Humphreys-lines
are identified. The dashed line indicates the Humphreys ionisation
edge.}
\end{center}
\end{figure}

}

\item{From 3.48\,\mic\ on, fringes at the end of band 1D affect the
ISO-SWS spectra of almost all stars in the sample.}

\item{A clear discrepancy is visible at the beginning of band 1A.
For the {\it{warm}} stars, the H5-22 and H5-23 lines emerge in
that part of the spectrum. An analogous discrepancy is also seen
for the {\it{cool}} stars, though it is somewhat more difficult to
recognise due to the presence of many CO features (Fig.\
\ref{dis1a}). Being present in the continuum of both {\it{warm}}
and {\it{cool}} stars, this discrepancy is attributed to problems
with the Relative Spectral Response Function (RSRF). A broad-band
correction was already applied at the short-wavelength edge of
band 1A \citep{Vandenbussche2001clim.confE..97V}, but the problem
seems not to be fully removed. At the band edges, the system
response is always small. Since the data are divided by the RSRF,
a  small problem with the RSRF at these places may introduce a
pronounced error at the band edge.

\begin{figure}[h!]
\begin{center}
\resizebox{0.5\textwidth}{!}{\rotatebox{90}{\includegraphics{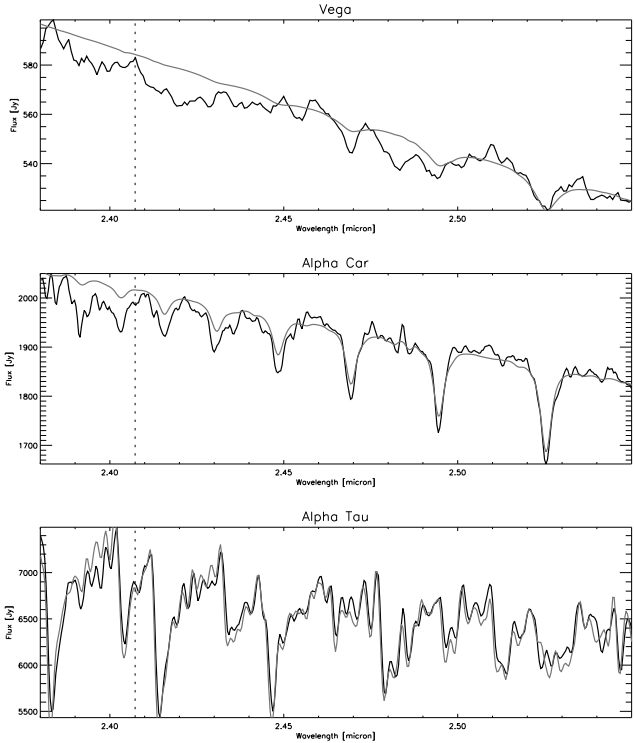}}}
\caption{\label{dis1a} Band 1A of Vega (A0~V), $\alpha$ Car (F0~II) and $\alpha$
Tau (K5~III) are displayed.  The ISO-SWS spectrum is plotted in black, the
synthetic spectrum in grey. A clear discrepancy is visible at the beginning of
this band ($\lambda$ $\lta 2.407$\,\mic).}
\end{center}
\end{figure}}

\item{Memory effects of the detectors make the calibration of band 2 for all the
stars very difficult. These memory effects are more severe for the
{\it{cool}} stars, since the CO and SiO absorptions cause a steep
increase (decrease) in flux for the up (down) scan (Fig.\
\ref{band2}). Band-border ratios are determined from the composite
SEDs by Cohen or from the synthetic spectrum. The RSRFs for the
sub-bands will therefore only be modelled well once there is a
fool-proof method to correct SWS data for detector memory effects.
Two other examples of this problematic RSRF modelling are visible
in band 2B and band 2C. For almost all {\it{warm}} and {\it{cool}}
stars, there is a `dip' around 6\,\mic\ and a jump in flux level
around 9.3\,\mic\ (Fig.\ \ref{band2}). This latter problem is a
residue of the correction for an instrumental absorption feature,
documented by \citet{Vandenbussche2001clim.confE..97V}. The
short-wavelength part of band 2A (where the CO $\Delta v = 1$
absorption starts) and of band 2C (with SiO $\Delta v = 1$) are
thus useless for passing a quantitative judgement upon the
parameters influencing the observed spectrum in this wavelength
range, like \teff, $\varepsilon$(C), $\varepsilon$(O), ... From
the longer wavelength parts of these same bands, one will not be
able to estimate directly fundamental stellar parameters, but
these data enable us to see if there are no contradictions between
this part of the ISO-SWS spectrum and the synthetic spectrum
generated using the parameters adopted from literature or
estimated from the ISO-SWS data in band 1.

\begin{figure}[h!]
\begin{center}
\resizebox{0.5\textwidth}{!}{\rotatebox{90}{\includegraphics{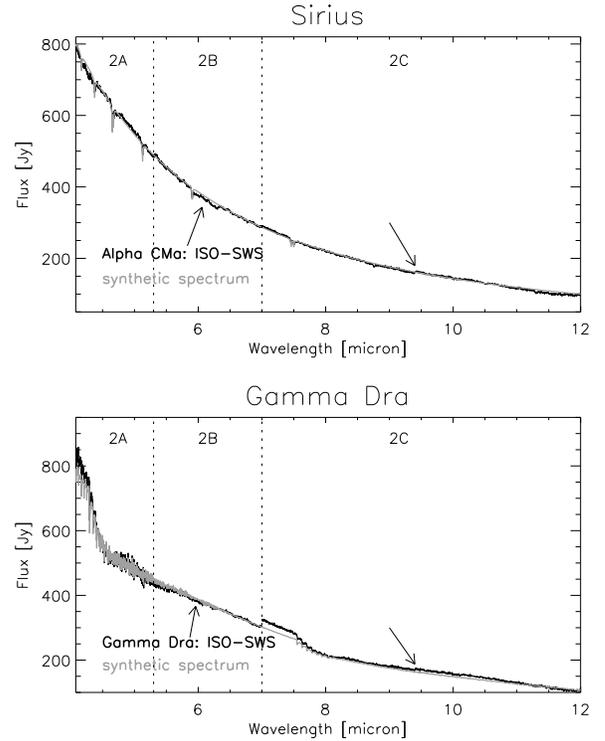}}}
\caption{\label{band2}Band 2 for Sirius (A1~V) and $\gamma$ Dra (K5~III) are
displayed. The ISO-SWS spectrum is plotted in black, the
synthetic spectrum in grey. Problems with the RSRF are indicated by
an arrow. A coloured version of this plot is available in the Appendix
as Fig.\ \ref{band2col}.}
\end{center}
\end{figure}
}

\end{enumerate}


\subsection{Cool stars: G2-M2}\label{cool}

\begin{enumerate}
\item{The situation changes completely when going to the {\it{cool}} stars of
the sample. While the spectrum of the {\it{warm}} stars is
dominated by atomic-line features, molecules determine the
spectral signature of the {\it{cool}} stars.  A few of the
--- problematic --- atomic features (see Sect.\ \ref{hot}) can still be
identified in these cool stars, e.g.\ the Mg-Si-Al-Ti-Fe spectral
feature around 3.97\,\mic\ (Fig. \ref{atomic}) remains visible for
the whole sample, even in $\beta$ Peg.}

\item{One of the most prominent molecular features in band 1 is
the first-overtone band of carbon monoxide (CO, $\Delta v = 2$)
around 2.4\,\mic. Already from $\alpha$ Cen A on, CO lines emerge
in band 1A, although the atomic features are still more dominant
for this star. It is striking that for all stars in the
sample with spectral type later than G2, the strongest CO features
(= band heads of $^{12}{\rm{CO}}$ 5-3, 6-4, 7-5, 8-6, and of
$^{13}{\rm{CO}}$  4-2, 5-3, 6-4 and 7-5)
are always predicted too strong compared to the ISO-SWS
observation (see, e.g., band 1A of $\gamma$ Dra in Fig.
\ref{restoohigh}, where both SWS and theoretical data are
rebinned to a resolving power of 1500, being the most conservative
theoretical resolution for band 1A \citep{Leech2002}). The only
exception is seen for the band head
of $^{13}{\rm{CO}}$ 4-2 since the problem with the RSRF at the
beginning of band 1A (see point 5.\ in previous section) dominates in
this wavelength range.

\begin{figure}[h!]
\begin{center}
\resizebox{0.5\textwidth}{!}{\rotatebox{90}{\includegraphics{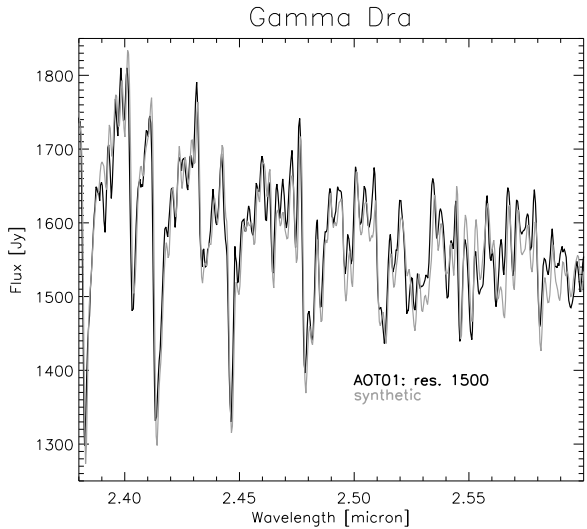}}}
\caption{\label{restoohigh}Comparison between the ISO-SWS observation and the
synthetic spectrum of $\gamma$ Dra at a resolving power of 1500 \citep[being
the most conservative theoretical resolution for band
1A][]{Leech2002}. A coloured version of this plot is available in the
Appendix as Fig.\ \ref{restoohighcol}.}
\end{center}
\end{figure}

In these oxygen-rich stars, the CO lines are a direct measure of
the C abundance. From the present spectra, this carbon
abundance can be estimated in two ways:
\begin{itemize}
\item{the strength of the CO spectral features is directly related
to $\varepsilon$(C);}
\item{the presence of many CO absorption lines causes a global
depression in the spectrum
up to $\sim$ 2.9\,\mic\ (see Fig.\ 4 in Paper~I), which may be used to determine
$\varepsilon$(C).}
\end{itemize}
Computing a synthetic spectrum with the carbon abundance
determined from this last criterion, results however in the
(strongest) computed CO spectral features being too strong
compared to the ISO-SWS observation, most visible in the strong
bandheads (2 -- 4\,\%). It has to be noted that this mismatch
occurs in band 1A, where the standard deviation of the rebinned
spectrum is larger than for the other sub-bands (see Fig.\ 6 in
\citetalias{Decin2000A&A...364..137D}) and that the error is
within the quoted accuracy of ISO-SWS in band 1A
\citep{Leech2002}. However, this mismatch is not random, in the
sense that the observed CO features are {\it{always}} weaker than
the synthetic ones.


A first step towards solving this problem was the use of the
high-resolution FTS-KP spectrum of $\alpha$ Boo (see Section
\ref{highres}). As was explained in Sect.\ \ref{highres} a
high-resolution synthetic spectrum was generated for $\alpha$ Boo
and rebinned to a resolution of 60000. The agreement between the
observed FTS-KP and synthetic spectrum is extremely good! One
example was already shown in Fig. \ref{FTS}, another one is
depicted in Fig.\ \ref{COfts} on page \pageref{COfts}. In this
latter figure, the problematic subtraction of the atmospheric
contribution causes the spurious features around 2.46700\,\mic\
for the summer FTS-KP spectrum and around 2.46735\,\mic\ for the
winter FTS-KP spectrum.{\footnote{The colour plots showing the
comparison between FTS-KP and synthetic spectrum for the
wavelength range 2.38 -- 4.08\,\mic, may be obtained at
www.ster.kuleuven.ac.be/\~\,leen/artikels/ISO2/FTS/. Spectral
ranges, for which the atmospheric contribution causes too much
trouble, are omitted.}} When scrutinising carefully the
first-overtone CO lines in the FTS-KP spectrum, it is obvious that
all the $^{12}$CO 2-0 lines, and almost all the $^{12}$CO 3-1
lines, are predicted as too {\it{weak}} (by 1 -- 2\,\%) and not as
too {\it strong} as was suggested in previous paragraph from the
comparison between ISO-SWS and synthetic spectra! Also the
fundamental CO lines are predicted as being a few percent too
weak. For a better judgement of the KP-SWS-synthetic
correspondence, the FTS-KP spectrum was rebinned to the ISO
resolving power. This was not so straightforward due to the
presence of the spurious features originating from the problematic
subtraction of the atmospheric contribution. In order to conserve
the flux, these spurious features were replaced by the flux value
of the synthetic spectrum. This adaptation is acceptable for the
following reasons:
\begin{itemize}
\item{most of the changes were made in regions that are not
strongly contributing to the most prominent features;}
\item{if for a particular feature the first argument does not
apply, the change mentioned above does still not introduce
additional errors. The FTS-KP-spectrum will be locally shifted
towards the ISO spectrum, since the synthetic predictions are (in
this particular case) systematically in between the ISO-SWS and FTS-KP
spectrum. Thus, if statistically relevant differences are still
detected, they will always correspond to an underestimate of the real
differences between the ISO-SWS and FTS-KP spectra.}
\end{itemize}


Fig.\ \ref{FTSISOsynCO} on page \pageref{FTSISOsynCO} shows the
comparison between the ISO-SWS,
the FTS-KP and the synthetic spectrum for $\alpha$ Boo at a resolving
power of 1500. It is clear that the differences between the
strength of the CO features in the ISO-SWS and FTS-KP spectrum are
significant. This is a strong indication for problems in the
{\it calibration} process. The question now arises where this
calibration problem originates from.

Firstly, it has to be noted that the flux values are unreliable in the
wavelength region from 2.38 to 2.40\,\mic\ due to problems with the
RSRF of band 1A; main features here include the CO 2-0 P18 and
the CO 2-0 P21 lines.

Secondly, no correlation is found with the local minima and maxima in the
RSRF of band 1A.

\begin{figure}[h!]
\begin{center}
\resizebox{0.5\textwidth}{!}{\rotatebox{90}{\includegraphics{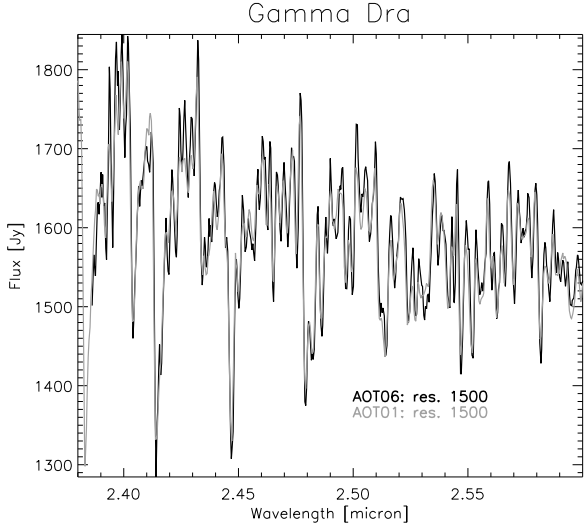}}}
\caption{\label{res1500}Comparison between the AOT06 observation
(revolution 538, black) and the AOT01 speed-4 observation (grey) of
$\gamma$ Dra. Both are rebinned to a resolving power of 1500. A
coloured version of this plot is available in the Appendix as Fig.\
\ref{res1500col}.}
\end{center}
\end{figure}

Defining the spectral resolution and instrumental profile for the ISO-SWS
grating spectrometers, is not straightforward \citep{Lorente1998,
Lutz1999}. Only for an AOT02 observation has the instrumental profile now been
derived quite accurately \citep{Lutz1999}. The AOT01 mode introduces an
additional smoothing which, due to the intricacies of SWS-data acquisition, is
different from a simple boxcar smooth. The simulation of SWS-AOT01 scans gives
non-gaussian profiles, as can be seen in Fig.\ 6 by \citet{Lorente1998}.
Nevertheless, the theoretical profile of a speed-4 observation
approximates a gaussian profile very closely. Therefore, since the instrumental
profile of an AOT01 is still not exactly known, the synthetic data were
convolved with a gaussian with FWHM=$\lambda$/resolution. This incorrect
gaussian instrumental profile introduces an error which will be most visible on
the strongest lines.

\begin{figure}[h!]
\begin{center}
\resizebox{0.5\textwidth}{!}{\rotatebox{90}{\includegraphics{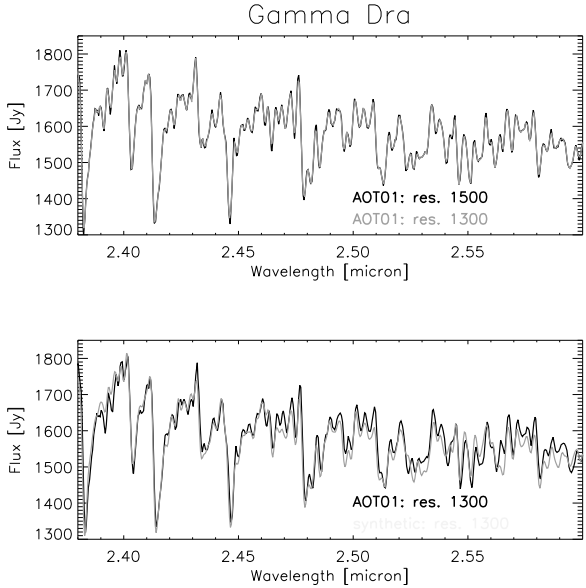}}}
\caption{\label{resolution}\underline{Top:} Comparison between the
AOT01 speed-4 observation of $\gamma$ Dra rebinned at 1) a
resolving power of 1500 (black) and 2) a resolving power of 1300 (grey).
\underline{Bottom:} Comparison between the ISO-SWS AOT01 observation
in band 1A (black) and its synthetic spectrum (grey) at a resolving
power of 1300. A coloured version of this plot is available in the
Appendix as Fig.\ \ref{resolutioncol}.}
\end{center}
\end{figure}

Only for $\gamma$ Dra an AOT06 observation, scanning this
wavelength range, was available in the ISO data-archive. The
comparison be\-tween the AOT01 speed-4 and AOT06 observation, both
rebinned to a resolving power of 1500, is an indication that the
resolving power of an AOT01 speed-4 observation in band 1A is lower
than 1500 (Fig.\ \ref{res1500}). The theoretical resolving power for an
AOT01 speed-4 observation in band 1A is $\ge 1500$. Fig.\ 5 in
\citet{Lorente1998} shows, however, a large deviation from this value,
attributed to the fainter continuum of the source used for
measuring the instrumental profile and the less accurate fitting in
this band. The use of a resolving power of 1300 --- instead of the
theoretical resolving power of 1500 --- for band 1A yields a) almost
no difference from the AOT01 SWS data at a resolving power of 1500 and b) a
better match between the SWS and synthetic data  for the
strongest CO features (Fig.\
\ref{resolution}). This observational resolving power of 1300 is also
in good agreement with the observational value given by \citet{Lorente1998} in
her Fig.\ 5.

The incorrect use of a gaussian instrumental profile, together
with too high a --- theoretical --- resolving power of 1500 form the origin
of the discrepancy seen for the strongest CO features. Therefore, the
resolving power of band 1A was taken to be 1300 instead of
1500. Part of the other discrepancies seen in band 1A may be
explained by problems with 1.\ the accuracy of the oscillator strengths of
atomic transitions in the near-IR (see point 1.\ in this section) or
2.\ the RSRF in the beginning of band 1A (see point 5.\ in previous section).}

\begin{figure}[h!]
\begin{center}
\resizebox{0.5\textwidth}{!}{\rotatebox{90}{\includegraphics{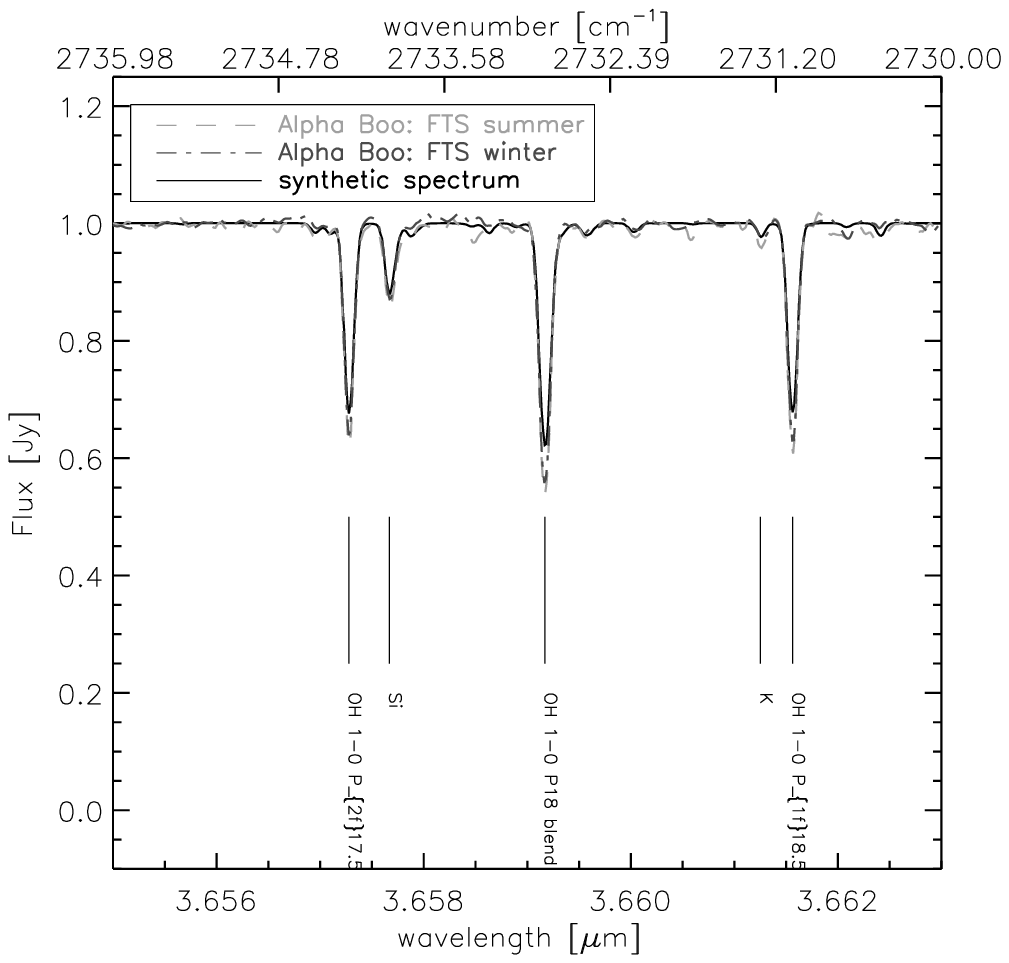}}}
\caption{\label{FTSOH} Summer and winter FTS-KP
spectra of $\alpha$ Boo at a resolving power $\lambda / \Delta
\lambda$ of 60000. They are compared with the synthetic spectrum
for $\alpha$ Boo with stellar parameters \teff\ = 4350\,K,
$\log$ g = 1.50, M = 0.75\,\Msun, [Fe/H] = $-0.50$, \vt\ = 1.7\,\kms, \cc\ =
7, $\varepsilon$(C) = 7.96, $\varepsilon$(N) = 7.61,
$\varepsilon$(O) = 8.68, $\varepsilon$(Mg) = 7.33 and $\Gamma_t$ =
3\,\kms. The OH 1-0 lines are predicted as too weak. A coloured
version of this plot is available in the Appendix as Fig.\
\ref{FTSOHcol}.}
\end{center}
\end{figure}

\item{Concentrating in the OH-lines, we see that for
both the high-re\-so\-lu\-tion FTS-KP (Fig.\ \ref{FTSOH}, Fig. \ref{FTS})
and the medium-resolution SWS spectra (Fig.\ \ref{FTSISOsynOH}), the
strongest lines (OH 1-0 and OH 2-1 lines) are
predicted as too weak, while the other OH lines match very well.
Since the same effect occurs for these two different observations,
it is plausible to assume that the origin of the problem is
situated in the theoretical model or in the synthetic-spectrum
computation. Wrong oscillator strengths for the OH lines could,
e.g., cause that kind of problems. Being based on different
electric dipole moment functions (EDMFs), the OH-line lists of
Sauval \citep{Melen1995}, \citet{Partridge+Schwenke1995} and
\citet{Goldman1998}, do however all show the same trend. The difference
in $\log gf$-value for the main branches is small ($< 0.01$)
between Goldman and Sauval, and is somewhat larger ($< 0.05)$
between Schwenke and Sauval. Thus, no systematic error seems to
occur in the oscillator strenghts.  Since a similar discrepancy
was also noted for the low-excitation CO lines, some assumptions on
which the models are based, e.g.\ homogeneity, hydrostatic
equilibrium, are cleary questionable for these stars.

}

\item{From 3.48\,\mic\ on, fringes at the end of band 1D affect the
ISO-SWS spectra of almost all stars in the sample.}

\item{Also for the {\it{cool}} stars, the same remark as given in point
6 for the {\it{warm}} stars concerning the memory effects applies. Clearly,
the fundamental bands of CO and SiO are present for almost all
{\it{cool}} stars, though no accurate quantitative scientific
interpretation can be made due to these memory effects.}

\end{enumerate}

\section{Implications on calibration and modelling}\label{impact}

The results of this detailed comparison between observed ISO-SWS data and
synthetic spectra have an impact both on the calibration of the ISO-SWS data and
on the theoretical description of stellar atmospheres.

From the calibration point of view, a first conclusion is reached
that the broad-band shape of the relative spectral response
function is at the moment already quite accurate, although some
improvements can be made at the beginning of band 1A (see Fig.\
\ref{vglbroadrsrf}) and band 2. Also, a fringe pattern is
recognised at the end of band 1D. Inaccuracies in the adopted
instrumental profile used to convolve the synthetic spectra with,
together with too high a resolving power, may cause the strongest
CO $\Delta v = 2$ band heads in the observed ISO-SWS spectrum to
be weaker than the predicted line strength.

\begin{figure}[h]
\begin{center}
\resizebox{0.5\textwidth}{!}{\rotatebox{90}{\includegraphics{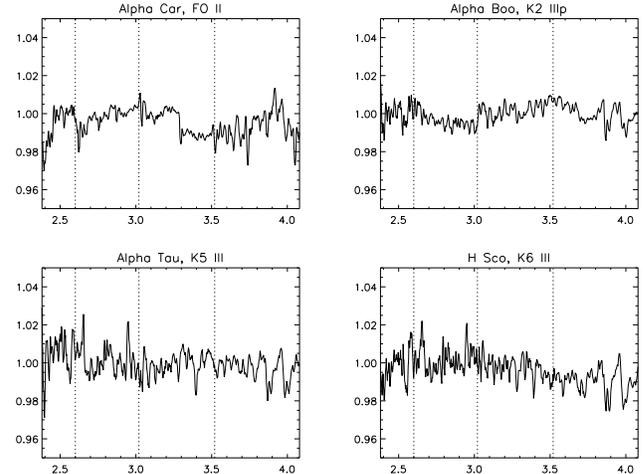}}}
\caption{\label{vglbroadrsrf}The observed ISO-SWS spectra of some
stars with different spectral types are divided by their synthetic
spectrum calculated by using the atomic line list of
\citet{Hirhor95}. The ratio is rebinned to a resolving power of
250. Only
at the beginning of band 1A the same trend is visible in all stars,
indicating that the broad-band shape of the relative
spectral-response functions is already quite accurately known for
the different sub-bands of band 1, with the only exception being
at the beginning of band 1A. The feature arising around 3.85\,\mic\
for the cool stars is an atomic feature discussed is Section \ref{hot}.}
\end{center}
\end{figure}

\begin{figure}[t]
\begin{center}
\resizebox{0.5\textwidth}{!}{\rotatebox{90}{\includegraphics{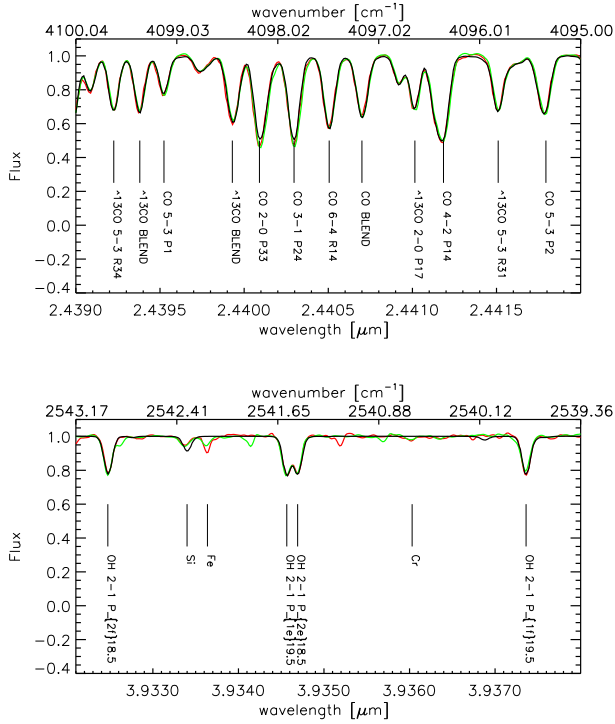}}}
\vspace*{-5ex}
\caption{\label{FTS} Summer (red) and winter (green) FTS-KP spectrum of
$\alpha$ Boo at a resolving power of 60000.  The differences between
the summer and winter FTS-KP spectrum of $\alpha$ Boo are mainly the
result of noise resulting from the removal of the telluric
spectrum.  Due to the opposite heliocentric shifts of the spectra, the
spectrum of Arcturus is fully recovered in spite of the telluric
lines.  The Arcturus spectra are compared with the synthetic spectrum
(black) for $\alpha$ Boo with stellar parameters \teff\ = 4350\,K, $\log$ g =
1.50, M = 0.75\,\Msun, [Fe/H] = $-0.50$, \vt\ = 1.7\,\kms, \cc\ = 7,
$\varepsilon$(C) = 7.96, $\varepsilon$(N) = 7.61, $\varepsilon$(O)
= 8.68, $\varepsilon$(Mg) = 7.33, $\varepsilon$(Si) = 7.20 and
$\Gamma_t$ = 3\,\kms.}
\end{center}
\end{figure}

\begin{figure}[h!]
\begin{center}
\resizebox{0.47\textwidth}{!}{\rotatebox{90}{\includegraphics{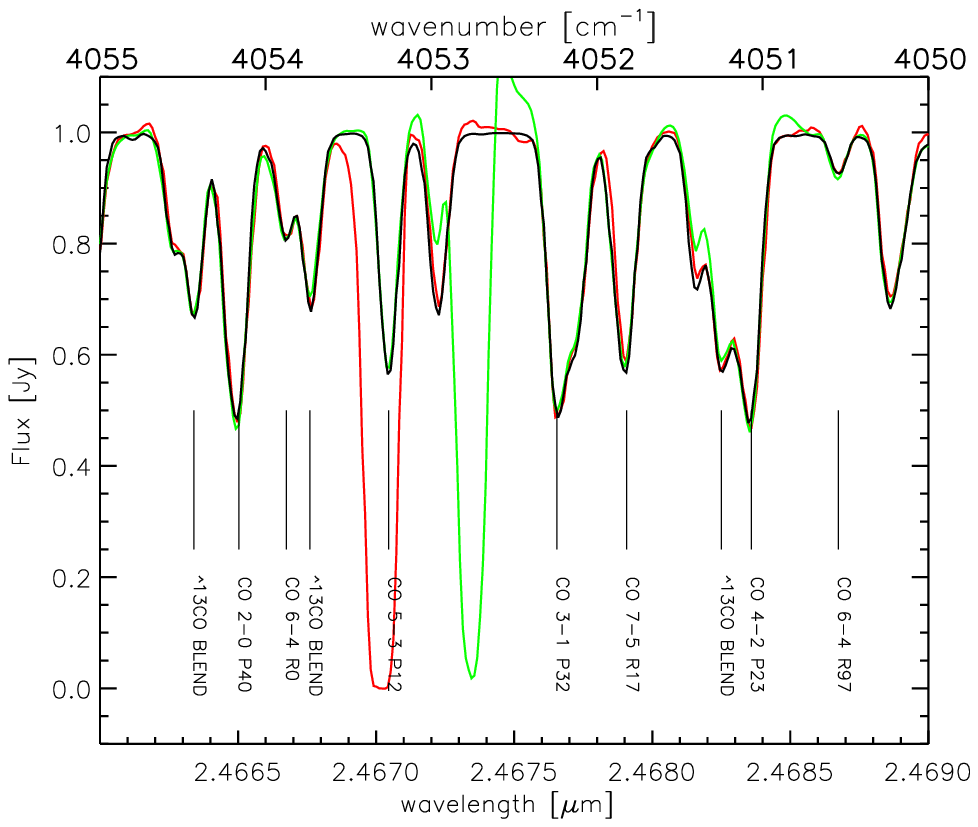}}}
\vspace{-5ex}
\caption{\label{COfts}Summer (red) and winter (green) FTS-KP spectra
of $\alpha$ Boo at a resolving power $\lambda / \Delta \lambda$ of
60000. They are compared with the synthetic spectrum (black) for
$\alpha$ Boo with stellar parameters \teff\ = 4350\,K, $\log$ g =
1.50, M = 0.75\,\Msun, [Fe/H] = $-0.50$, \vt\ = 1.7\,\kms, \cc\ = 7,
$\varepsilon$(C) = 7.96, $\varepsilon$(N) = 7.61, $\varepsilon$(O)
= 8.68, $\varepsilon$(Mg) = 7.33 and $\Gamma_t$ = 3\,\kms.}
\end{center}
\end{figure}

\begin{figure}[t]
\begin{center}
\resizebox{0.5\textwidth}{!}{\rotatebox{90}{\includegraphics{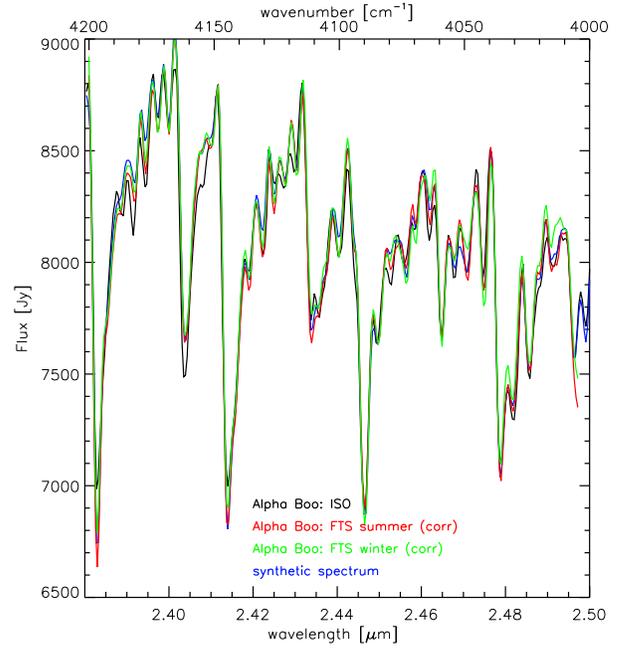}}}
\caption{\label{FTSISOsynCO}Comparison between CO spectral
features of the ISO-SWS spectrum of $\alpha$ Boo (black), the
summer FTS-KP spectrum (red), the winter FTS-KP spectrum (green) and the
synthetic spectrum (blue) at a resolving power of 1500.}
\end{center}
\end{figure}

\begin{figure}[b]
\begin{center}
\resizebox{0.5\textwidth}{!}{\rotatebox{90}{\includegraphics{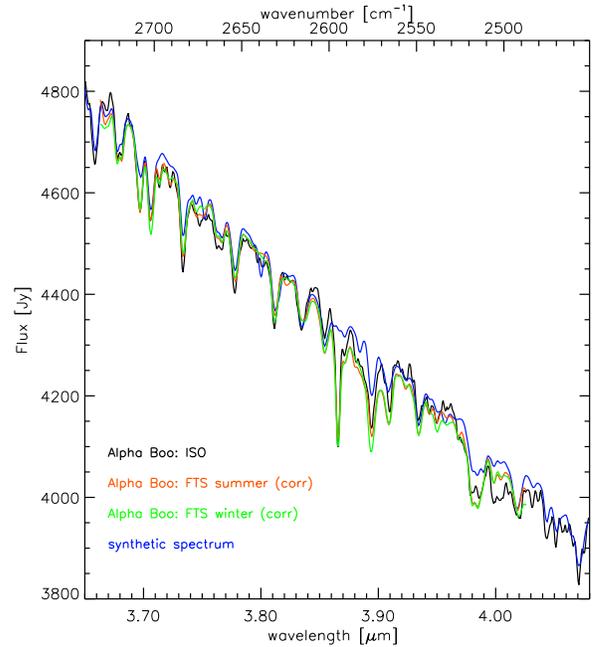}}}
\caption{\label{FTSISOsynOH}Comparison between OH spectral
features of the ISO-SWS spectrum of $\alpha$ Boo (black), the
summer FTS-KP spectrum (red), the winter FTS-KP spectrum (green) and the
synthetic spectrum (blue) at a resolving power of 1500. The strong
atomic feature around 3.87\,\mic\ is discussed in Sect.\ \ref{hot}.}
\end{center}
\end{figure}

Concerning the modelling part, problems with the construction of
the theoretical model and the computation of the synthetic spectra are
pointed out. The comparison between the high-resolution FTS-KP
spectrum of $\alpha$ Boo and the corresponding synthetic spectrum
revealed that the low-exci\-ta\-tion first-overtone (and fundamental)
CO lines and fundamental OH lines are predicted as being a few percent too weak,
indicating that some assumptions, on which the models are based, are
questionable for cool stars. This is not surprising: in the
MARCS-code radiative equilibrium is assumed, also
for the outermost layers, implying that a temperature bifurcation, caused by
e.g.\ effects of convection and convective overshoot with
inhomogeneities in the upper photosphere, can not be allowed
for. Consequently, the cores of e.g.\ CO lines --- or the saturated CO
lines in general --- are not described with full success.
Noting however the high level of accordance between observations
and theoretical predictions for many molecular lines, we may conclude
that the oscillator strengths for these molecular transitions are now
already accurate enough in order to use these lines to test some of
the assumptions made in the mathematical stellar atmosphere code:
e.g.\ the temperature distribution can be
disturbed in order to simulate a chromosphere, convection, a
change in opacity, ... 
The complex computation of the hydrogen lines, together with the
inaccurate atomic oscillator strengths in the infrared rendered
the computation of the synthetic spectra for {\it{warm}} stars
difficult. Therefore, one of us (J.\ S.)
has derived empirical oscillator strengths from the
high-resolution FTS-ATMOS spectrum of the Sun \citep[for more details, we
refer to Paper~V of this series and to][]{Sauval2000}.

Although it was impossible to perform a detailed comparison
between observed and synthetic data in band 2, we may conclude
that the continua of the synthetic spectra in this wavelength
range are reliable, but that strong molecular lines (e.g. band
heads of CO $\Delta v = 1$ and SiO $\Delta v = 1$) may be
predicted a few percent too weak. Nevertheless, combining the
synthetic spectra of both warm and cool stars, allowed us to test
the recently developed method for memory effect correction
\citep{Kester2001clim.confE..46K} and to re-derive the relative
spectral response function for bands 1 and 2 for OLP10
\citep{Vandenbussche2001clim.confE..97V}. The results of these
tests will be described in Paper~V of this series (Decin et al.,
2002, in preparation). In conjunction with photometric data, this
same input data-set was used for the re-calibration of the
absolute flux-level of the spectra observed with ISO-SWS
\citep{Shipman2001clim.confE..18S}. In this way, both consistency
and integrity were implemented. Moreover, the synthetic spectra of
the standard sources of our sample are not only used to improve
the flux calibration of the observations taken during the nominal
phase, but they are also an excellent tool to characterise
instabilities of the SWS spectrometers during the post-helium
mission.

Although we have mainly concentrated on the discrepancies between
the ISO-SWS and synthetic spectra --- since this was the main task
of this research --- we would like to emphasise the good agreement
between observed ISO-SWS data and theoretical spectra. The small
discrepancies still remaining in band 1 are at the 1 -- 2\,\%
level for the giants, proving not only that the calibration of the
(high-flux) sources has already reached a good level of accuracy,
but also that the description of cool star atmospheres and
molecular line lists is quite accurate.

\begin{acknowledgements}LD acknowledges support from the Science Foundation of
Flanders.  This research has made use of the SIMBAD database, operated
at CDS, Strasbourg, France and of the VALD database, operated at Vienna,
Austria. It is a pleasure to thank the referees, J.\ Hron and F.\
Kupka, for their careful reading of the manuscript and for their
valuable suggestions.
\end{acknowledgements}


\aareferences

\end{document}